\documentclass[useAMS,usedcolumn,usenatbib]{./reference/mn2e}

\usepackage{epsfig,graphicx,natbib}
\usepackage{./reference/mycite}

\begin{document}
\title[Photo-z from the ground and weak lensing from space]{Photometric redshifts for weak lensing tomography from space: the role of optical and near infra-red photometry}
\author[]
{F. B. Abdalla$^{1}$,
A. Amara$^{2}$,
P. Capak$^{3}$,
E. S. Cypriano$^{1}$,
O. Lahav$^{1}$,
J. Rhodes$^{3,4}$\\
$^{1}$Department of Physics and Astronomy, University College London,
Gower Street, London, WC1E 6BT, UK.\\
$^{2}$Service d\'Astrophysique, CEA Saclay, 91191 Gif sur Yvette, France.\\
$^{3}$California Institute of Technology, 1201 E California Blvd., Pasadena, CA 91125, USA.\\
$^{4}$Jet Propulsion Laboratory, 4800 Oak Grove Drive, Pasadena, CA 91109, USA.
}
\maketitle

\begin{abstract}
We study in detail the photometric redshift requirements needed for
tomographic weak gravitational lensing in order to measure accurately
the Dark Energy equation of state.
In particular, we examine how ground-based photometry (u,g,r,i,z,y)
can be complemented by space-based near-infrared (IR) photometry
(J,H), e.g. on board the planned DUNE satellite. Using realistic
photometric redshift simulations and an artificial neural network
photo-z method we evaluate the Figure of Merit for the Dark Energy
parameters $(w_0, w_a)$. We consider a DUNE-like broad optical filter
supplemented with ground-based multi-band optical data from surveys
like the Dark Energy Survey, Pan-STARRS and LSST.  We show that the
Dark Energy Figure of Merit would be improved by a factor of 1.3 to 1.7
if IR filters are added on board DUNE.  Furthermore we show that with
IR data catastrophic photo-z outliers can be removed effectively.
There is an interplay between the choice of filters, the magnitude
limits and the removal of outliers.  We draw attention to the
dependence of the results on the galaxy formation scenarios encoded
into the mock galaxies, e.g the galaxy reddening.  For
example, very deep u band data could be as effective as the IR.  We also
find that about $10^5-10^6$ spectroscopic redshifts are needed for
calibration of the full survey.

\end{abstract}


\section{Introduction.}




\begin{figure*}
\begin{center}
\includegraphics[width=8.5cm,angle=0]{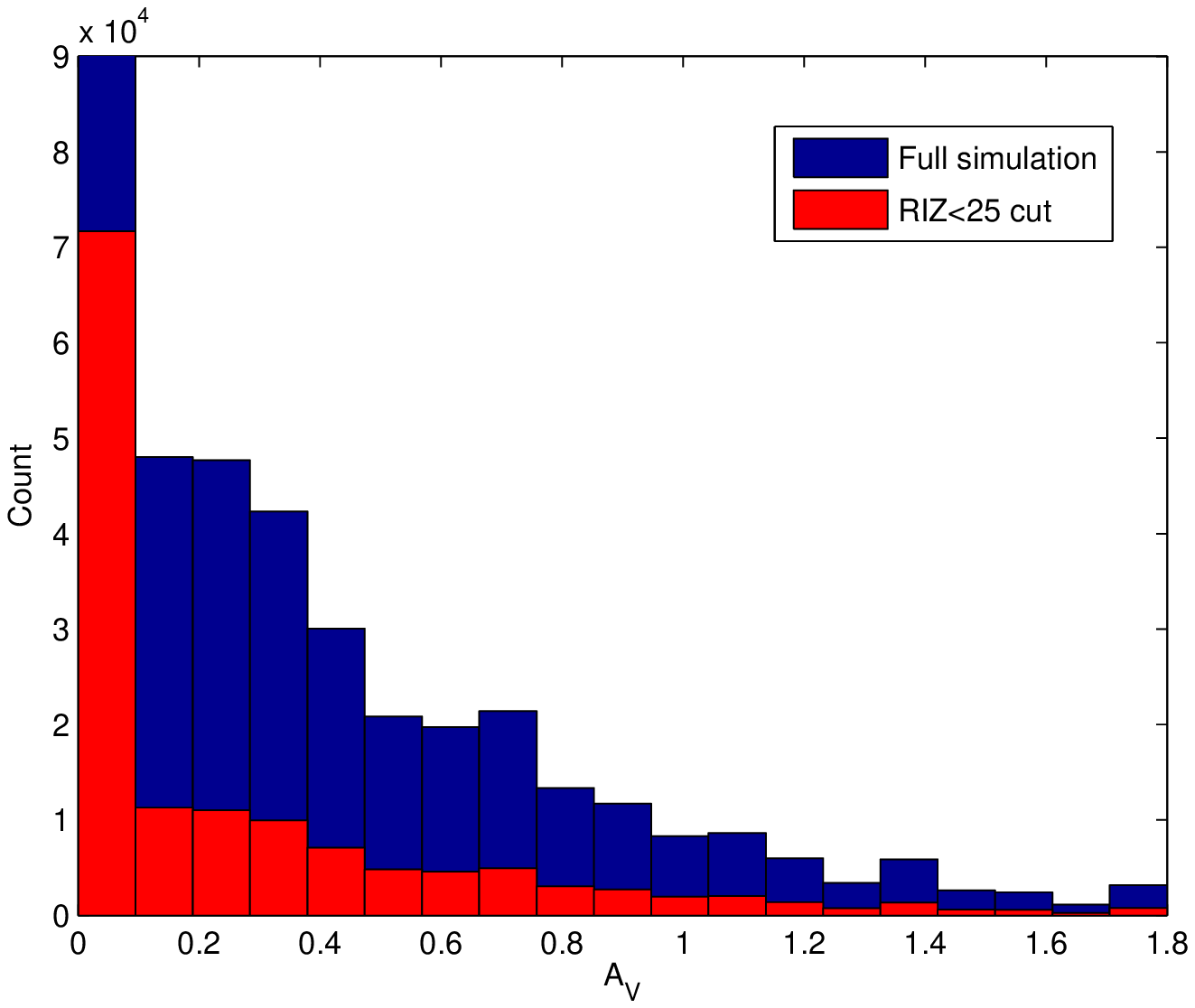}
\includegraphics[width=8.5cm,angle=0]{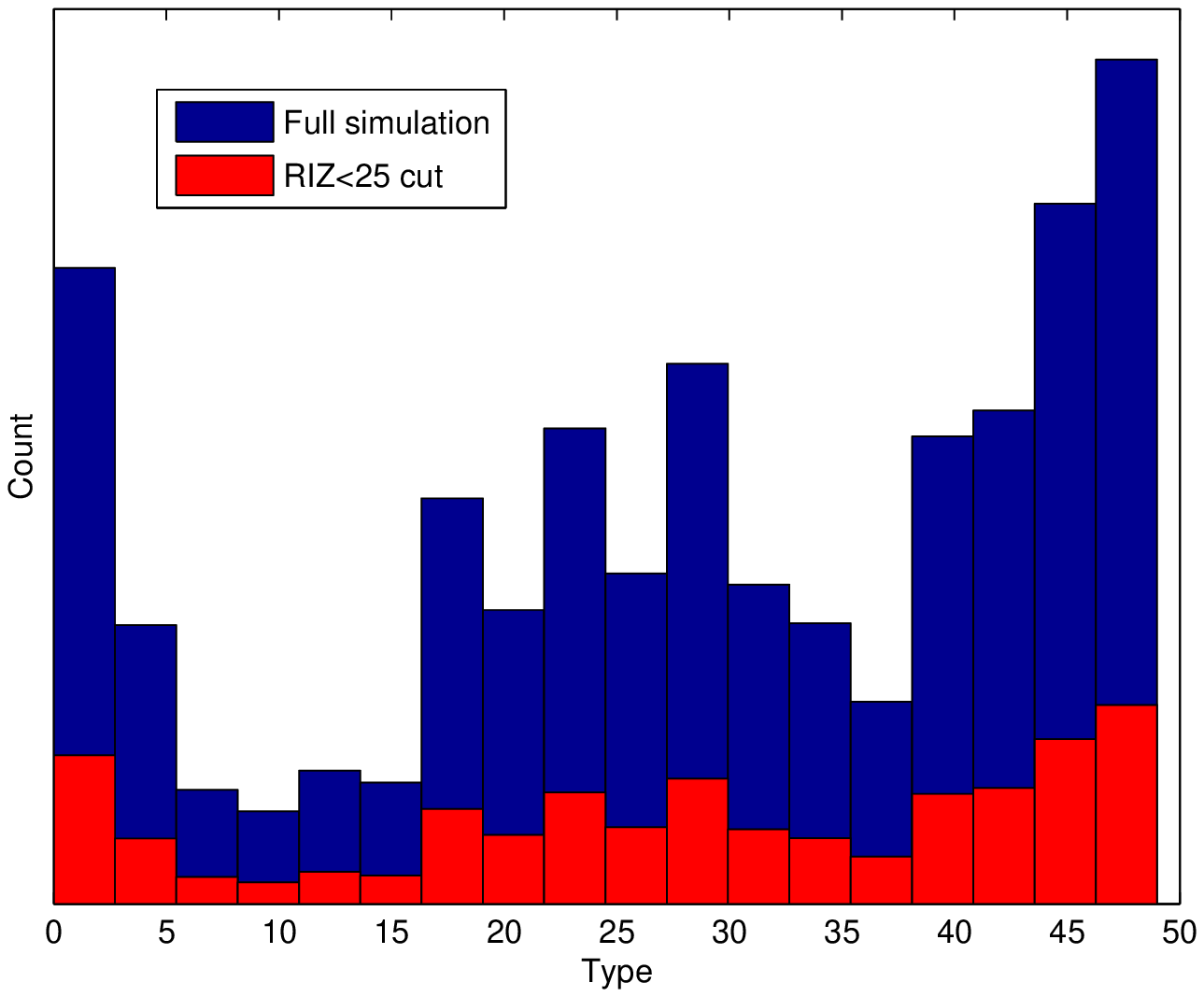}
\includegraphics[width=8.5cm,angle=0]{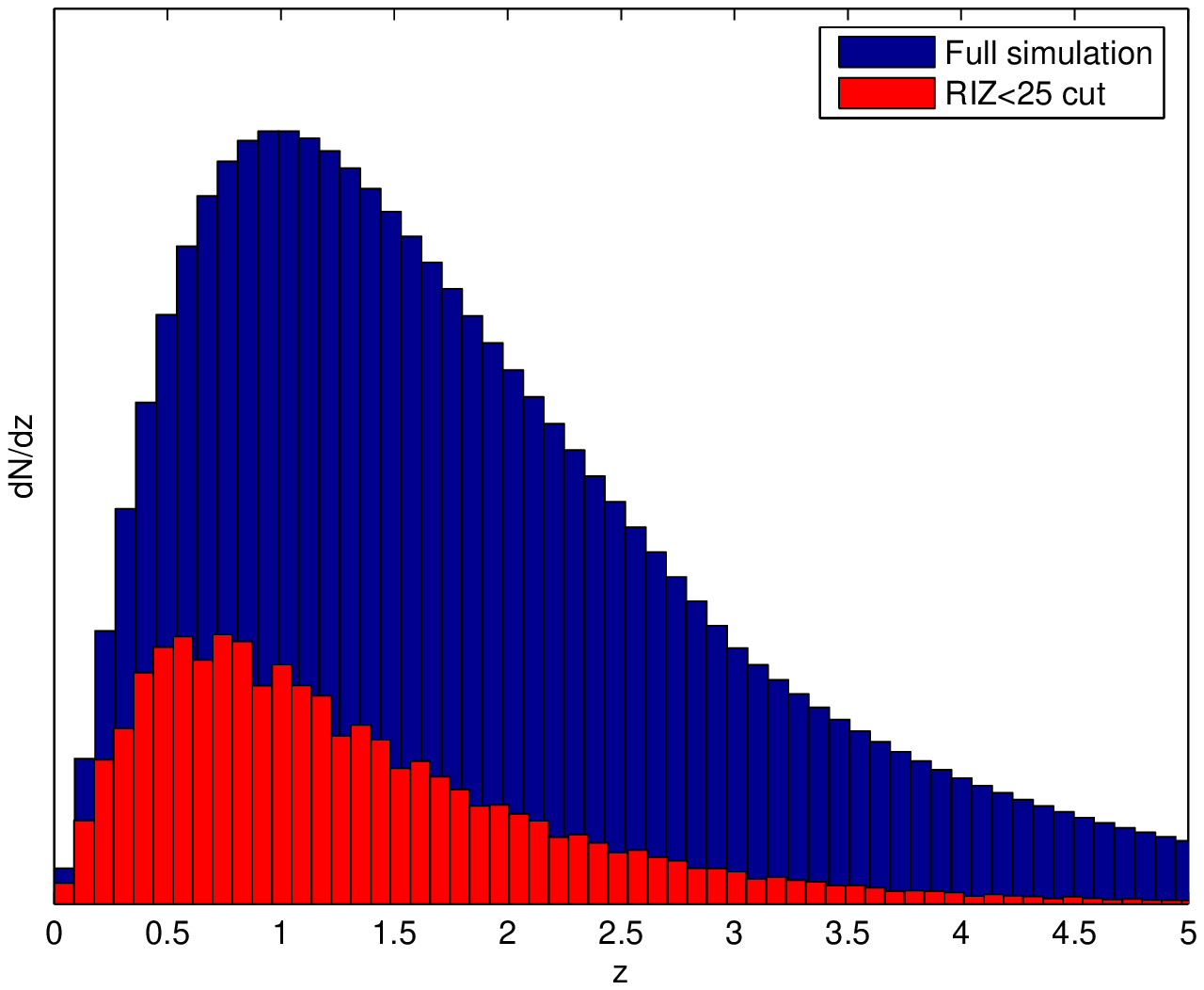}
\caption{The distributions characterising our simulations.
The number density of objects as a function of redshift for the full simulation,
which has a magnitude limit of 27, as well as the cut simulations with
a magnitude limit in the RIZ band of 25. We also plot the relative number
of galaxy types in the simulation with templates numbered from 0 to 50
according to Sec.\ref{sec::cat}, as well
as the amount of reddening ($A_v$, the extinction in magnitudes in V band)
applied to templates in the simulation.
\label{fig:cat}}
\end{center}
\end{figure*}



Measuring the nature of Dark Energy has become a central part of current
studies in cosmology.  A variety of methods have the potential to
probe Dark Energy through its effects on both structure growth and
geometry of the Universe
\citep{1997PhRvD..56.4439T,1998AJ....116.1009R,1999ApJ...517..565P,1999ApJ...522L..21H,2003ApJ...594..665B,2003ApJ...598..720S,2003PhRvD..68f3004H,2004NewAR..48.1063B}.  Future cosmic shear measurements are now widely believed to
have the greatest potential for constraining the Dark Energy equation
of state parameter \citep{2006Msngr.125...48P,2006astro.ph..9591A}.
However, to reach this potential future surveys will require tight
control over systematic and statistical errors.

Weak-lensing surveys will have to overcome systematic checks such
as a reliable point spread function (PSF)
removal, well calibrated shape estimation
\citep[e.g.][]{2006astro.ph..8643M},
removal of intrinsic galaxy alignments
\citep[e.g.][]{2002A&A...396..411K,2003A&A...398...23K,2003MNRAS.339..711H},
shear-shape alignment removal
\citep[e.g.][]{2004PhRvD..70f3526H,2006MNRAS.371..750H,2007ApJ...655L...1B} and
photometric redshifts bias corrections,
all of which are vital for the statistical
limits of the test to be achieved.

A reliable removal of PSF effects can be greatly improved by
using satellites such as the Dark Universe Explorer \citep[DUNE,
][]{2006SPIE.6265E..58R} or the Supernova/Acceleration Probe
\citep[SNAP, ][]{aldering-2004}, which will take advantage of the
image quality achievable in space to accurately measure the shapes
of lensed galaxies.
However, obtaining the multi-band photometry required to measure
photometric redshifts from space
would require a substantial mission and may not be necessary. At
certain wavelengths, obtaining photometry from the ground is much
faster. Given that large ground based facilities are available or
being planned the case for obtaining photometric redshifts from the
ground and shape measurements (and some photometry) form space is
compelling.

In this paper we analyse future weak lensing projects with space
based imaging for shape measurments plus ground  based multi-colour
photometry. We assess the impact of the photometric redshift
accuracy on the science goals of such a mission. We further assess
whether the science would benefit significantly by having some
photometry from space, especially very red optical and infrared
bands which are difficult to obtain from the ground due to the sky
brightness at these wavelenghts.

We begin by describing the catalogue generation upon which we base
our findings. We then make a full photometric redshift analysis
with artificial neural networks to assess the accuracy of the photo-z
obtained in each of the scenarios considered. We  explain the
different degeneracies found, their origin and how the different survey bands
and depths help in breaking those degeneracies. Finally we translate
our photometric redshift errors into a figure of merit for the
science considered,
in this case for the Dark Energy equation of state
derived via weak gravitational lensing tomography. All magnitudes stated
in this paper are AB magnitudes.

\section{Catalogue Generation.}
\label{sec::cat}

In order to simulate the catalogues used in this work, we used the
GOODS-N spectroscopic sample from \citet{2004AJ....127.3137C} and
\citet{2004AJ....127.3121W}. This data set includes a $R<24.5$
magnitude limited sample along with x-ray, radio, and colour
selected objects.  The majority of objects are at $z<1.4$. However,
a specific effort was made to include objects at $1.4<z<4$.  Using
these redshifts, we have generated a series of galaxy templates from
the broadband photometry, (U,B,V,R,I,Z,J,H,K,HK') using a method
similar to \citet{1999ASPC..191...19B}. The key difference is that
we assume a prior set of templates (\citet{1980ApJS...43..393C}
(CWW) + \citet{1996ApJ...467...38K} + intermediate types). We assign
the following types to templates: type 0 is Elliptical, type 10 is
Sbc, type 20 is Scd, type 30 is Irr, type 40 is SB3 and type 50 is
SB2.  Intermediate types are a linear interpolation of these types.
For instance, type 5 is 0.5El + 0.5Sbc. We, therefore, can remove
reddening from the photometry before constructing the templates.
Once we have a set of templates, the best fit SED and reddening
value for each object is found. We use the Calzetti reddening law
\citep{1997AJ....113..162C}. In addition to reddening, we apply a
correction for intergalactic absorption using the Madau law
\citep{1995ApJ...441...18M}.

We use the model described below for the luminosity function
evolution in the simulation to estimate the RIZ magnitude and
redshift distribution. The RIZ filter is assumed to be a broad
filter covering roughly the range 5500 \AA $\,$ to 10000 \AA. The
local r band luminosity function at $z=0$ is taken as well as the
\citet{1999ApJ...519....1S} luminosity function at $z=3$. We
linearly interpolate between them in redshift space.  At $z>3$ we
assume $L^\star$ and the faint end slope are constant, but the
volume density which decreases to $10^{-6} h^3 {\rm Mpc}^{-3}$ at $z
= 10$. We run a Monte-Carlo simulation which draws the RIZ magnitude
from this distribution and the reddening and spectral types from the
GOODS-N \citep{2004AJ....127.3137C} distribution, which should close
for a DUNE-like survey since the GOODS spectroscopic limit is
$R<24.5$. We note here that the mock is an extrapolation for
$R>24.5$ because GOODS is not complete for magnitudes fainter than
this level. We argue that this is not a large effect because the
DUNE magnitudes that we consider are not much fainter than these
magnitudes, hence there will only be a small extrapolation for
galaxies between  $R>24.5$ and  $RIZ<24.5$.
\begin{table*}
  \begin{center}
    \begin{tabular}{|l|c|c|c|c|c|c|c|}
      \hline        Band & DES & Pan-4 & LSST & Ideal & Ideal + u & DUNE & Cosmos  \\
      \hline
      $u$        &  -   & -    &  23.9 & -    & 26.1 & -    & 25.1 \\
      $g$        & 24.6 & 25.9 &  26.1 & 26.1 & 26.1 & -    & 25.3 \\
      $r$        & 24.1 & 25.6 &  27.4 & 26.1 & 26.1 & -    & 25.3 \\
      $i$        & 24.3 & 25.4 &  26.2 & 26.2 & 25.9 & -    & 25.0 \\
      $z$        & 23.9 & 23.9 &  25.1 & 25.5 & 25.5 & -    & 24.1 \\
      $y$        & -    & 22.3 &  24.3 & 25.0 & 25.0 & -    & - \\
      $B$        & -    & -    &  -    & -    &  -   & -    & 25.3 \\
      $V$        & -    & -    &  -    & -    &  -   & -    & 25.2 \\
      $RIZ$      & -    & -    &  -    & -    &  -   & 25.0 & - \\
      $F814$     & -    & -    &  -    & -    &  -   & -    & 25.4 \\
      $J$        & -    & -    &  -    & -    &  -   & 23.4 & - \\
      $H$        & -    & -    &  -    & -    &  -   & 23.2 & - \\
      $K$        & -    & -    &  -    & -    &  -   & -    & 20.2 \\
      \hline
    \end{tabular}    \vspace{2mm}
  \end{center}

  \begin{center}
    \begin{tabular}{|l|c|c|c|c|c|c|c|}
      \hline        Band & DES & Pan-4 & LSST & Ideal & Ideal + u & DUNE &
Cosmos  \\
      \hline
      $u$        &  -   & -    &  1.1 &  -   &  8.3 & -  &  3.3 \\
      $g$        &  2.3 &  7.7 &  9.2 &  9.2 &  9.2 & -  &  4.4 \\
      $r$        &  2.5 & 10.0 & 52.5 & 15.8 & 15.8 & -  &  7.6 \\
      $i$        &  6.4 & 17.7 & 37.0 & 37.0 & 28.0 & -  & 12.2 \\
      $z$        &  5.1 &  5.1 & 15.3 & 22.1 & 22.1 & -  &  6.1 \\
      $y$        & -    &  2.4 & 14.9 & 28.3 & 28.3 & -  & -    \\
      $B$        & -    & -    &  -    & -    &  -  & -    & 13.6 \\
      $V$        & -    & -    &  -    & -    &  -  & -    &  7.6 \\
      $RIZ$      & -    & -    &  -    & -    &  -  & 10.0 & -    \\
      $F814$     & -    & -    &  -    & -    &  -  & -    & 14.5 \\
      $J$        & -    & -    &  -    & -    &  -  & 15.6 & - \\
      $H$        & -    & -    &  -    & -    &  -  & 21.5 & - \\
      $K$        & -    & -    &  -    & -    &  -  & -    & 1.2 \\
      \hline
    \end{tabular}    \vspace{2mm}
  \end{center}
  \caption{The assumed surveys that we investigate in
this work. (Top table) The values quoted are 10 sigma magnitudes for
extended sources in the AB system.
We have taken assumed depths for proposed ground based future imaging surveys
and a possible IR survey from space. We also simulate a current ongoing survey
over a much smaller area to compare our results.
(Bottom table) Signal-to-noise ratios for a 25.0 RIZ Sbc galaxy for each of the
surveys/filters. \label{tab::depths}}
\end{table*}

We then calculate fluxes for the galaxy based on the redshift, SED
type, reddening and filter profiles, normalising to the RIZ
magnitude sampled.  Finally, we add Gaussian noise to the fluxes,
then estimate the magnitudes and errors from the fluxes with noise
in the same way as a photometry package would. We plot in
Fig.\ref{fig:cat} the statistical properties of the catalogue, i.e.
the galaxy distribution as a function of redshift, reddening and
type.

The final catalogue contains galaxies which are complete in the
simulation out to a magnitude limit of 27.0 in RIZ. However given
the surveys we are to simulate we cut the catalogue depending on the
photometry available. One of the purposes of this study is to assess
the impact that space based IR photometry will have on the estimate
of Dark Energy parameters. For this purpose we study a fiducial
survey which can be achieved by a satellite with a 1.2m mirror. With
a 1500s exposure this would reach depths in the IR of around 23.0
and would reach a depth in optical bands of around 25.0 in AB. These
depths would be feasible with a mission such as DUNE (see
Table.\ref{tab::depths}). However, this does not restrict our study
as it is applicable to any other wide field imager that would
provide similar data although this is a difficult task from the
ground. Hence, for an analysis of a DUNE like survey we obtain a cut
catalogue which has a $10 \sigma$ detection in the RIZ filter.
Unless we state otherwise we cut the catalogue at an RIZ magnitude
of 25. Corresponding photometry is available for other filters
obtained from the ground which may be more or less noisy than the
RIZ detection.

\section{Photometric redshifts}
\label{sec:photo_z}

\begin{figure*}
\begin{center}
\begin{minipage}[c]{1.00\textwidth}
\centering
\includegraphics[width=8.5cm,angle=0]{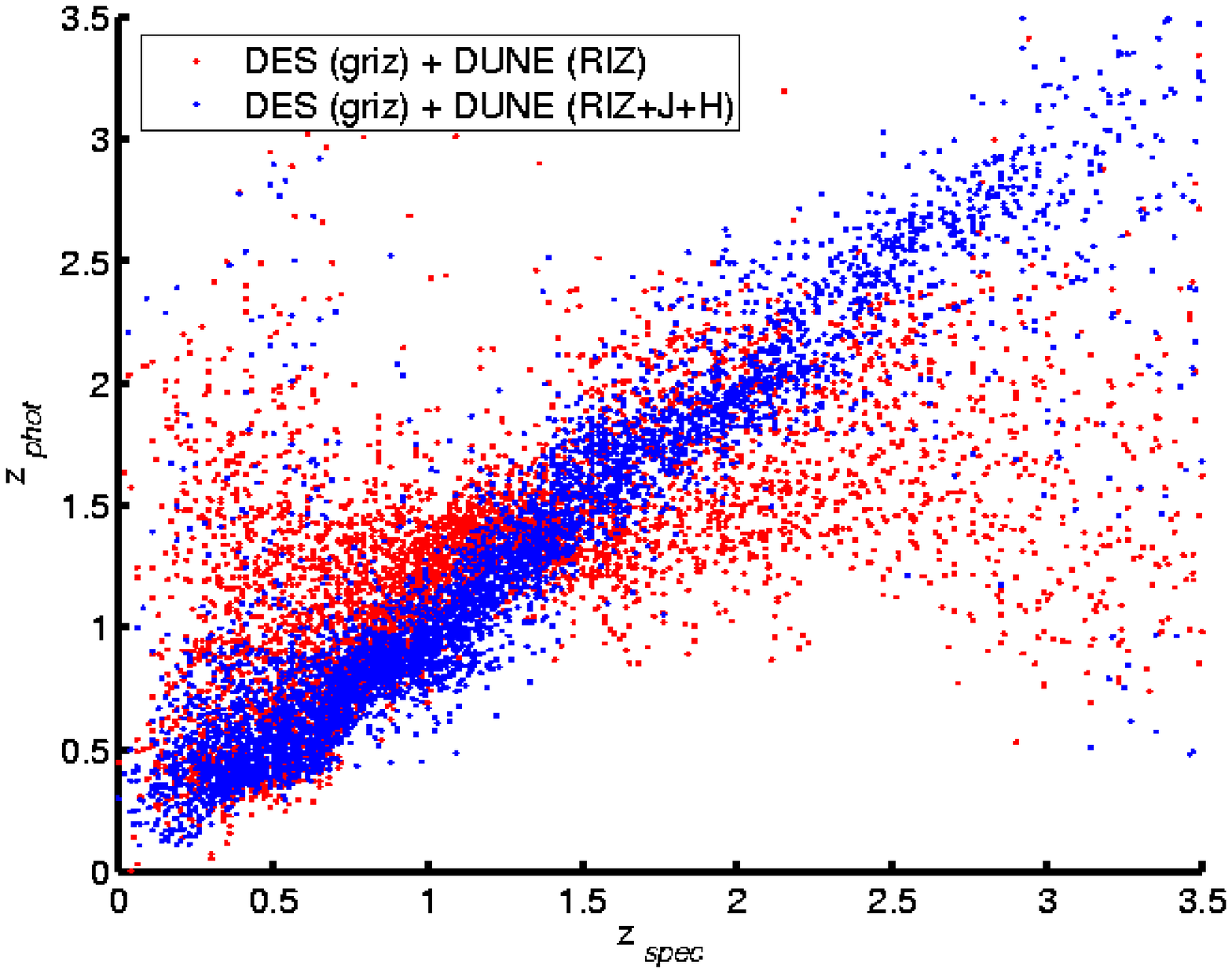}
\includegraphics[width=8.5cm,angle=0]{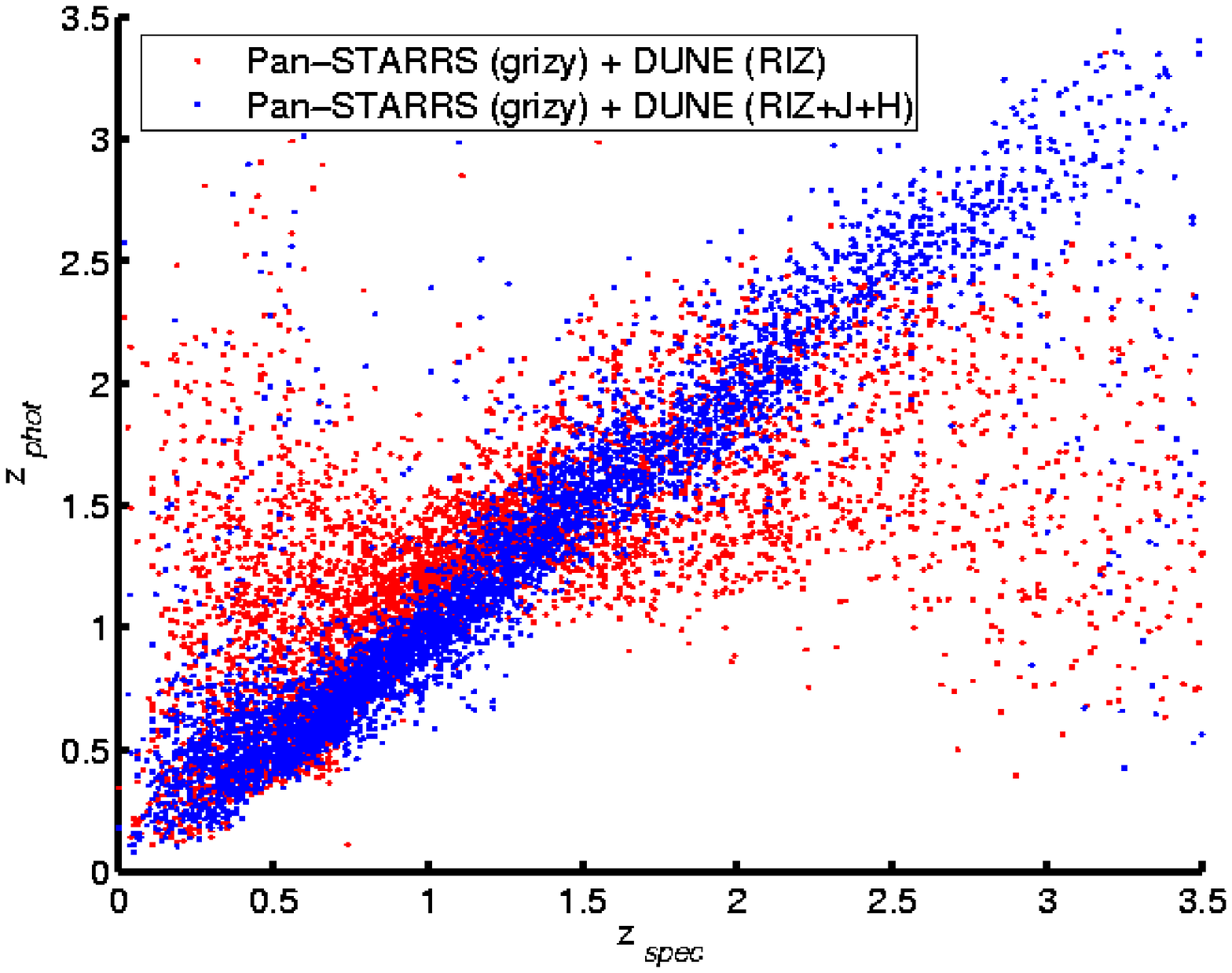}
\includegraphics[width=8.5cm,angle=0]{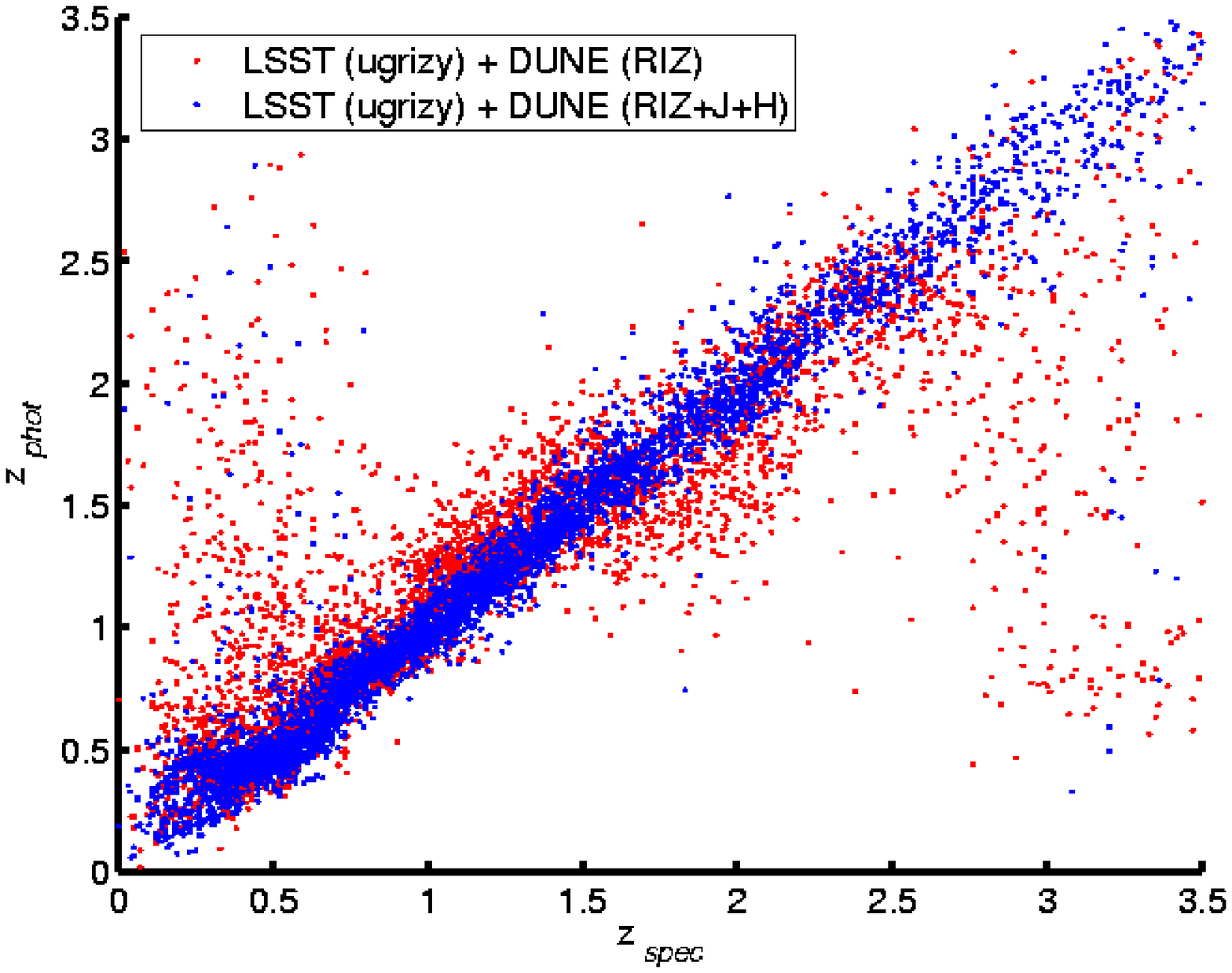}
\end{minipage}
\begin{minipage}[c]{1.00\textwidth}
\centering
\includegraphics[width=8.5cm,angle=0]{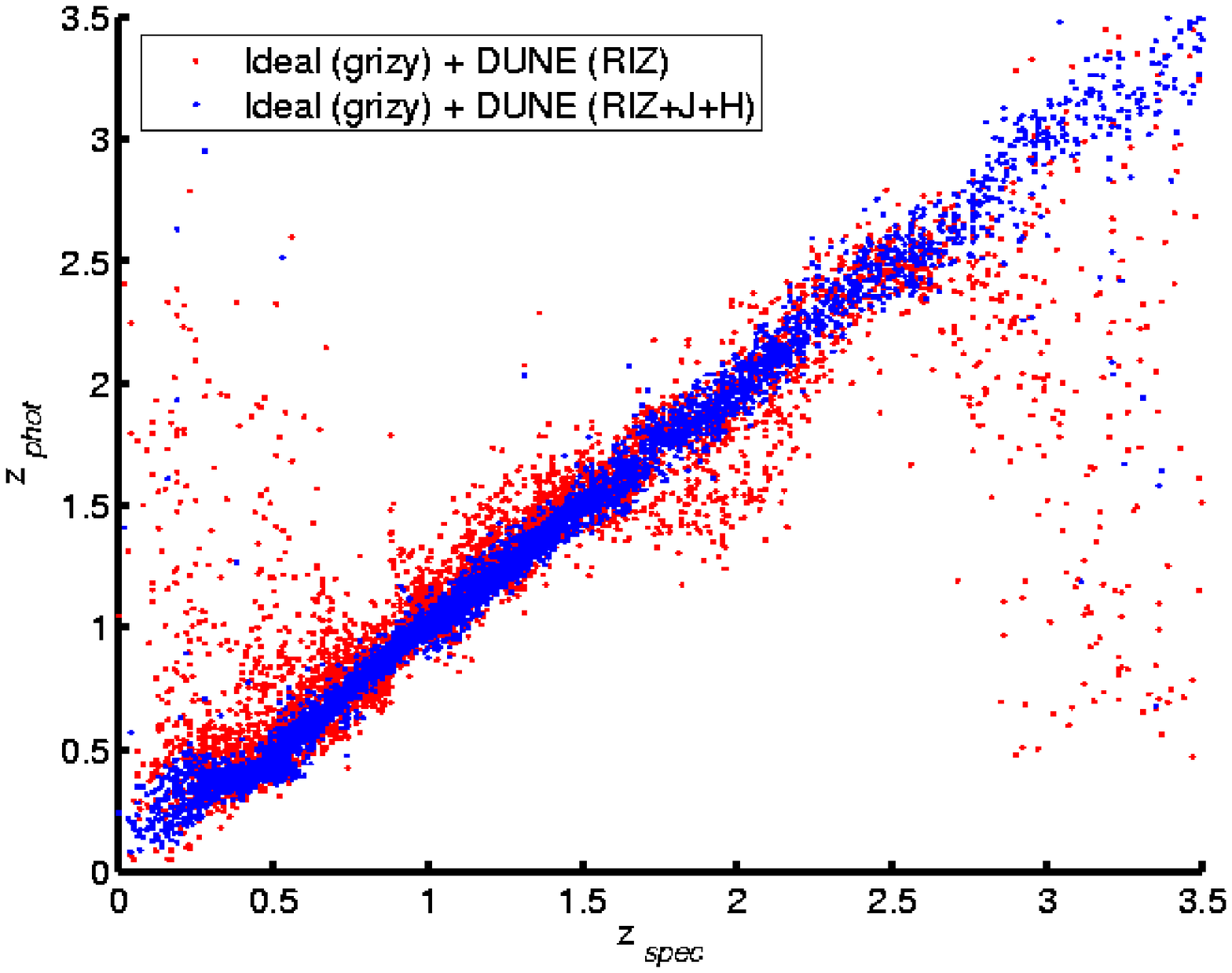}
\includegraphics[width=8.5cm,angle=0]{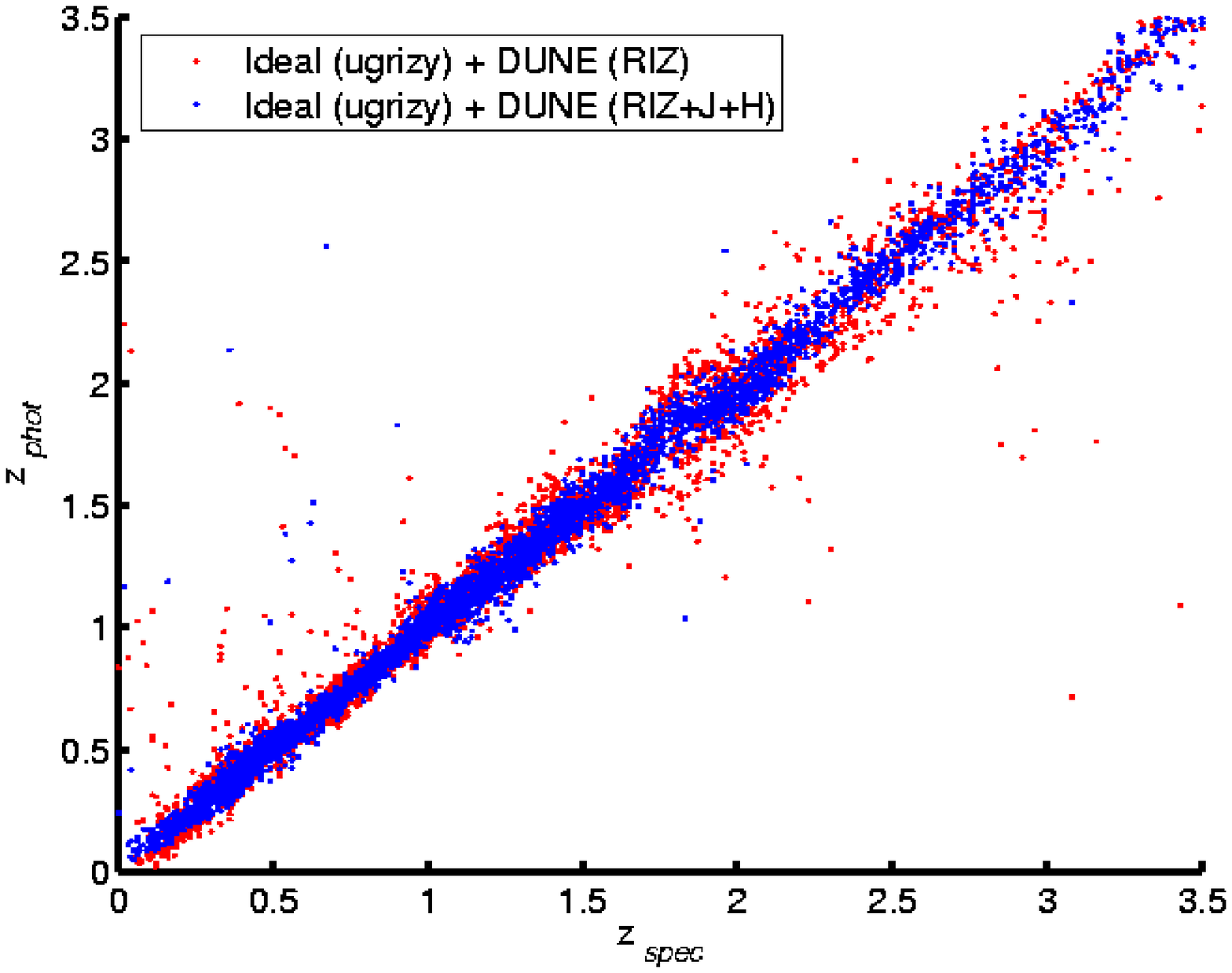}
\end{minipage}
\caption{Scatter plots of photometric redshifts
as a function of the true redshifts for some of the surveys
considered in Sec.\ref{sec:impact_bands}. We have shown the galaxies
which have photometric and spectroscopic redshifts below 3.5. We have
considered optical surveys with increasing depth.  The shallowest is  a survey with depth similar to DES, followed by a hypothetical
survey with depth similar to Pan-STARRS, and finally LSST. We also consider
two hypothetical optical surveys with very deep exposures in optical bands,
especially in the very red bands, one of which has a very deep u band exposure.
We assess in this figure how the inclusion of deep IR data obtained
from space would enhance the photometric redshift estimation.
\label{fig:zin_zout_IR}}
\end{center}
\end{figure*}

\begin{figure*}
\begin{center}
\begin{minipage}[c]{1.00\textwidth}
\centering
\includegraphics[width=8.8cm,angle=0]{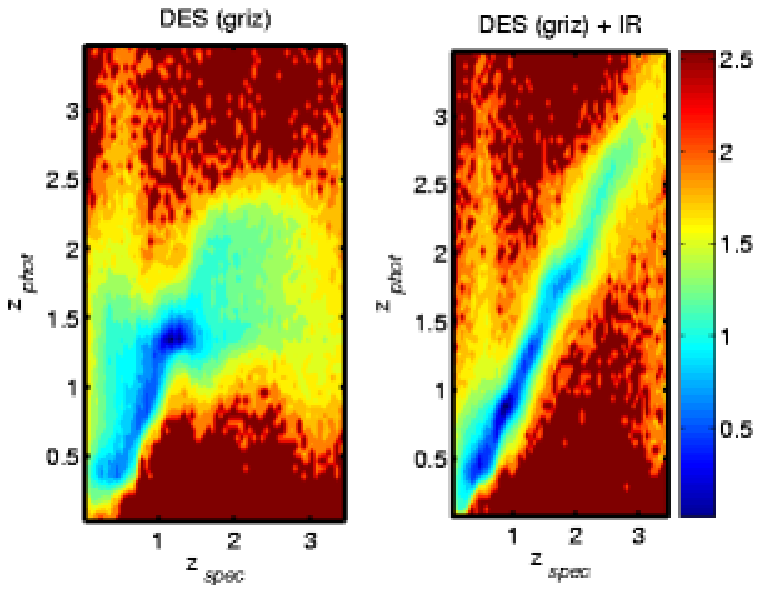}
\includegraphics[width=8.8cm,angle=0]{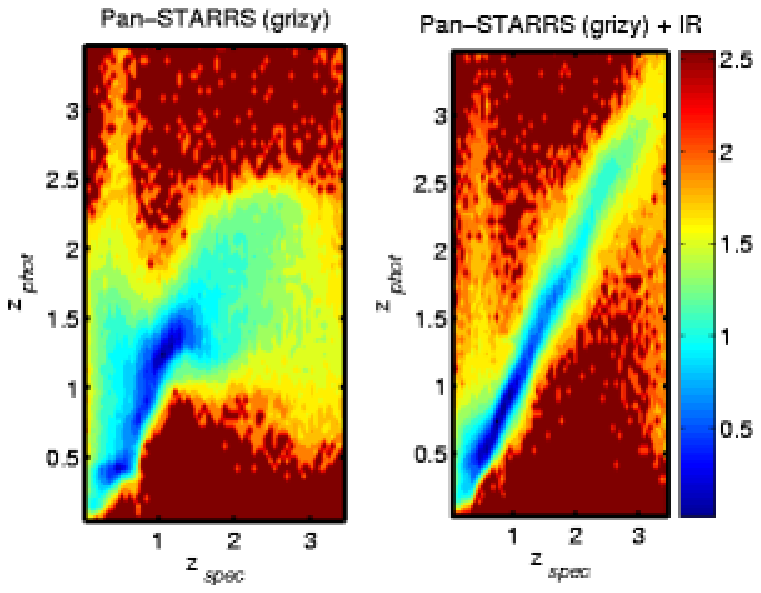}
\includegraphics[width=8.8cm,angle=0]{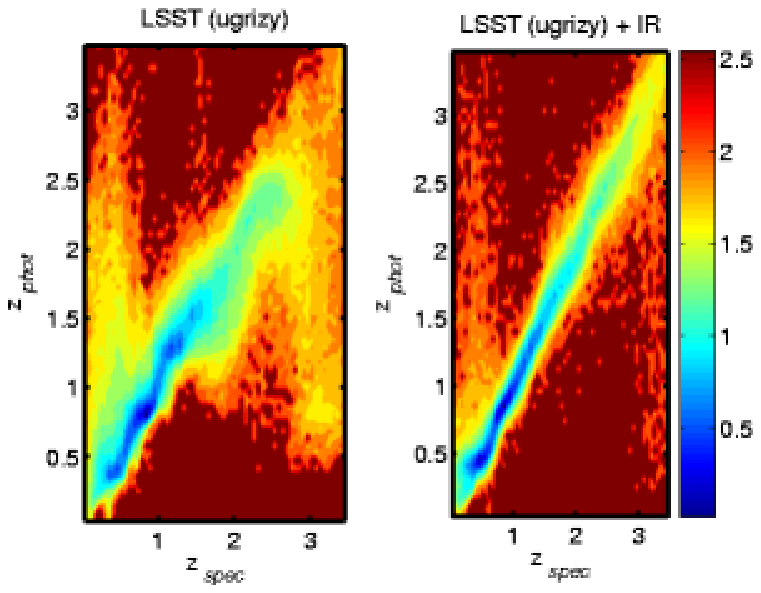}
\end{minipage}
\begin{minipage}[c]{1.00\textwidth}
\centering
\includegraphics[width=8.8cm,angle=0]{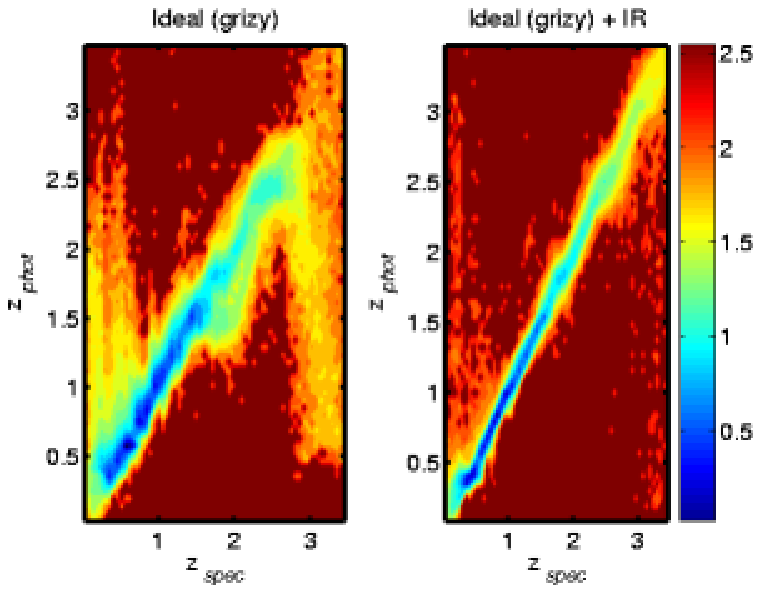}
\includegraphics[width=8.8cm,angle=0]{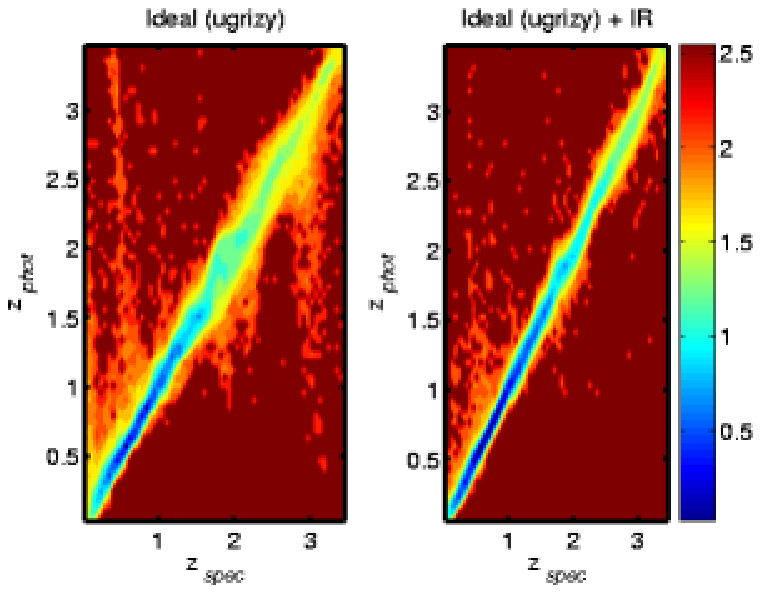}
\end{minipage}
\caption{Density maps in the
$z_{spec} - z_{phot}$ plane. An alternate representation of the data shown in
Fig.\ref{fig:zin_zout_IR}. The figures are colour coded according to
the local density of points. The colour scheme is exponential; this means that
a colour difference which is different by one unit according to the scale
means that the density is a factor of $e \simeq 2.718$ smaller.
We have shown the galaxies
which have photometric and spectroscopic redshifts below 3.5 but the
neural network fit has been done over the entire range of galaxies available.
\label{fig:zin_zout_IR_dens}}
\end{center}
\end{figure*}

\subsection{Estimating photo-z via neural networks}
\label{sec:photo_z_annz}

Essentially
there are two approaches to obtain reliable
photometric redshifts. Template methods compare the colours found with the
photometric data for each galaxy with the colours that templates would
predict were these templates placed at different redshifts. Training
methods attempt to map out a function which would translate magnitudes
to a single photometric redshift. A more detailed description
of methods can be found in \citet{2003AJ....125..580C} and references within.

It is well studied and accepted that template methods are more versatile
and can be applicable when no spectroscopic data is available. However,
methods that use training set methods are more reliable and produce
better photometric redshifts.

In this work we use artificial neural networks (ANNz),
a training set method which has been shown to produce
competitive results compared to other training set methods available
\citep{2004PASP..116..345C}.
ANNz is a supervised neural network training tool. It requires a training set
which is the data  used to optimise the cost function

\begin{equation}
E=\sum_k(z_{\mathrm{phot}}(\mathbf{c},\mathbf{m}_{k}) -
z_{\mathrm{train},k})^2, \label{eqn.cost}
\end{equation}

\noindent with respect to the free parameters (`weights'), $\mathbf{c}$, where
the sum is over the
galaxies in the training set which determines
the goodness of fit of the training set and $\mathbf{m}_{k}$ are the
magnitudes of each galaxy . If the data is noisy, a validation
set is also required in order to prevent over-fitting. This is another portion
of the data which also has spectroscopic information available but that is
not included in the training process. It is solely used to provide the
error function to be minimised.

\begin{figure*}
\begin{center}
\includegraphics[width=8.5cm,angle=0]{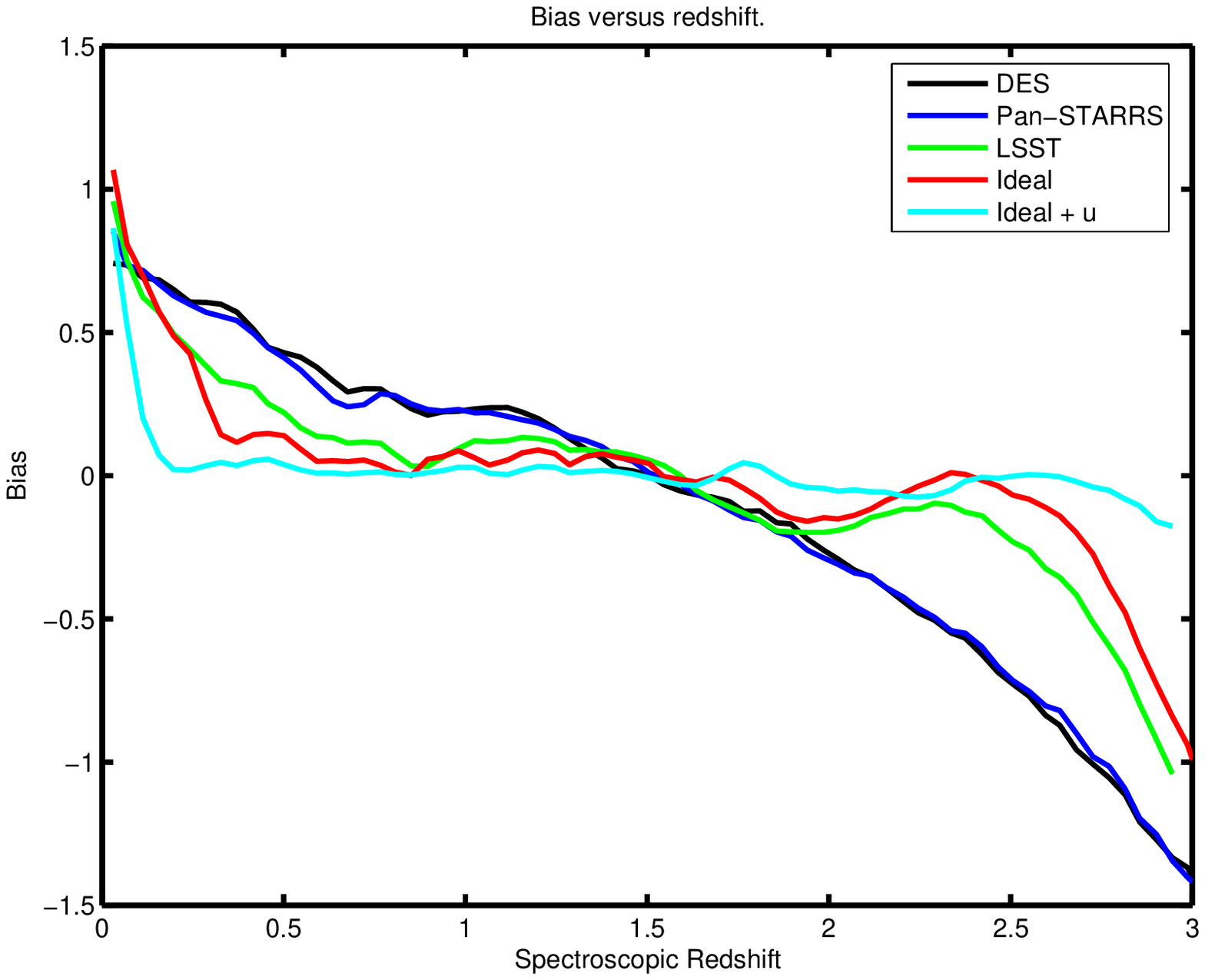}
\includegraphics[width=8.5cm,angle=0]{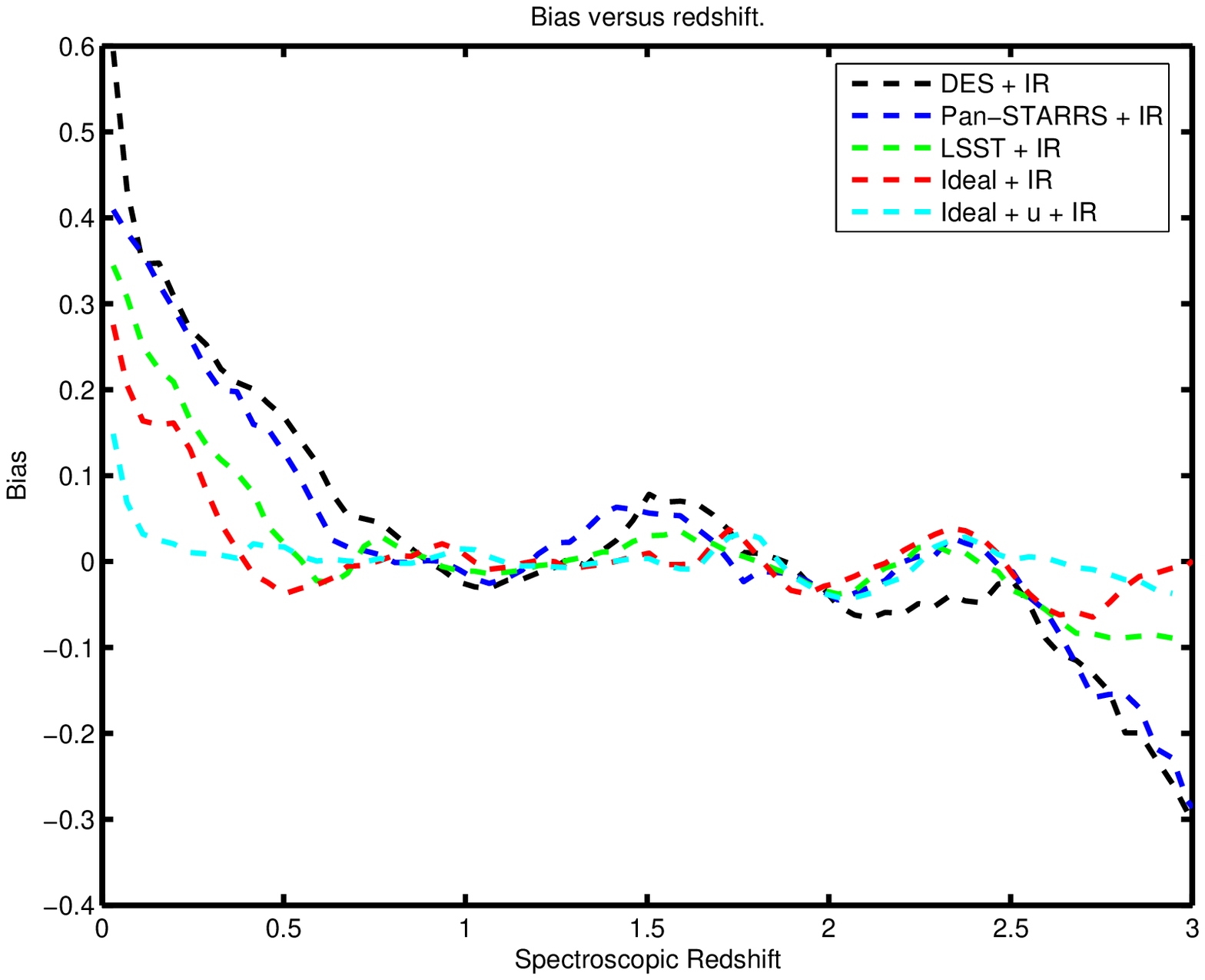}

\includegraphics[width=8.5cm,angle=0]{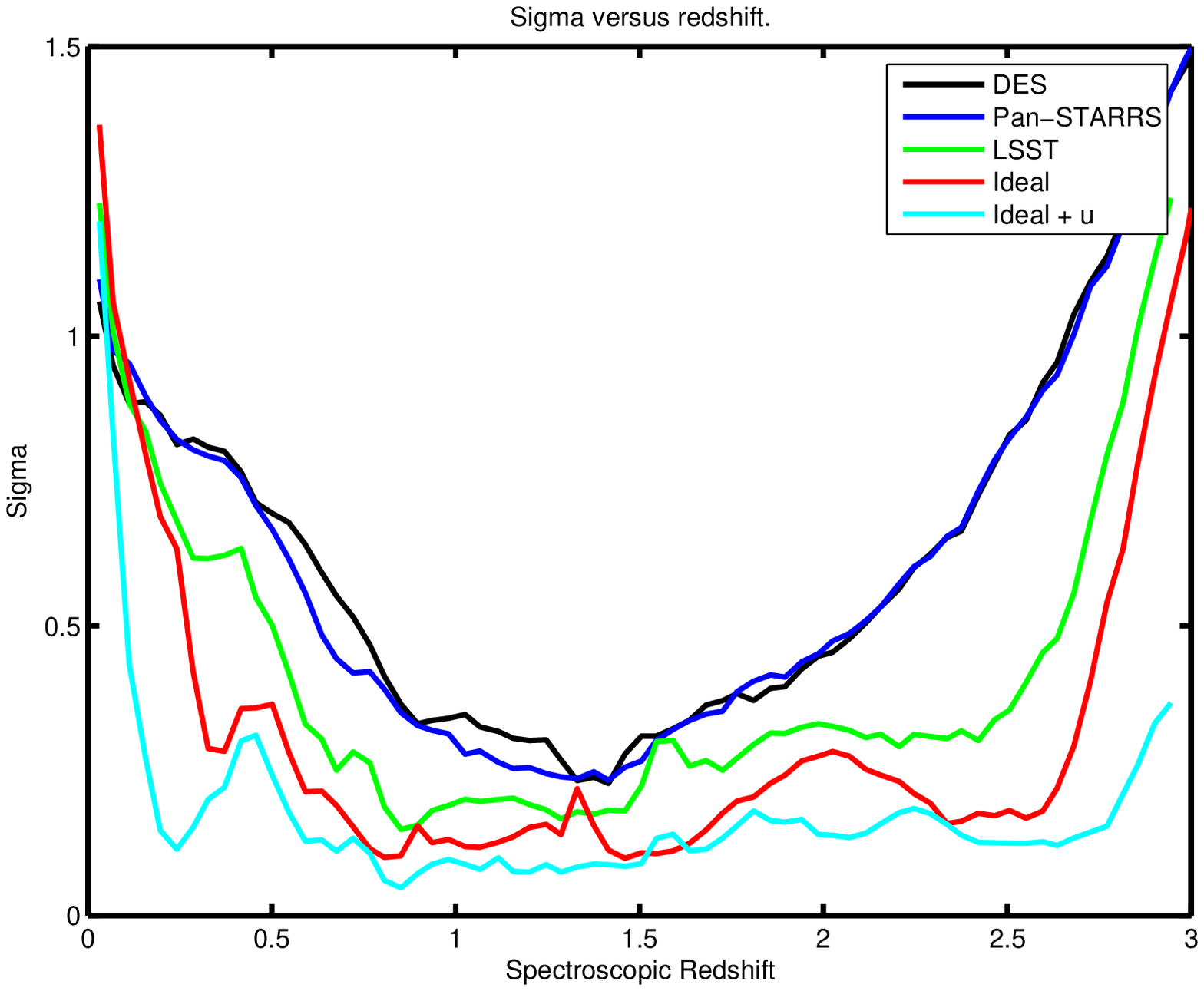}
\includegraphics[width=8.5cm,angle=0]{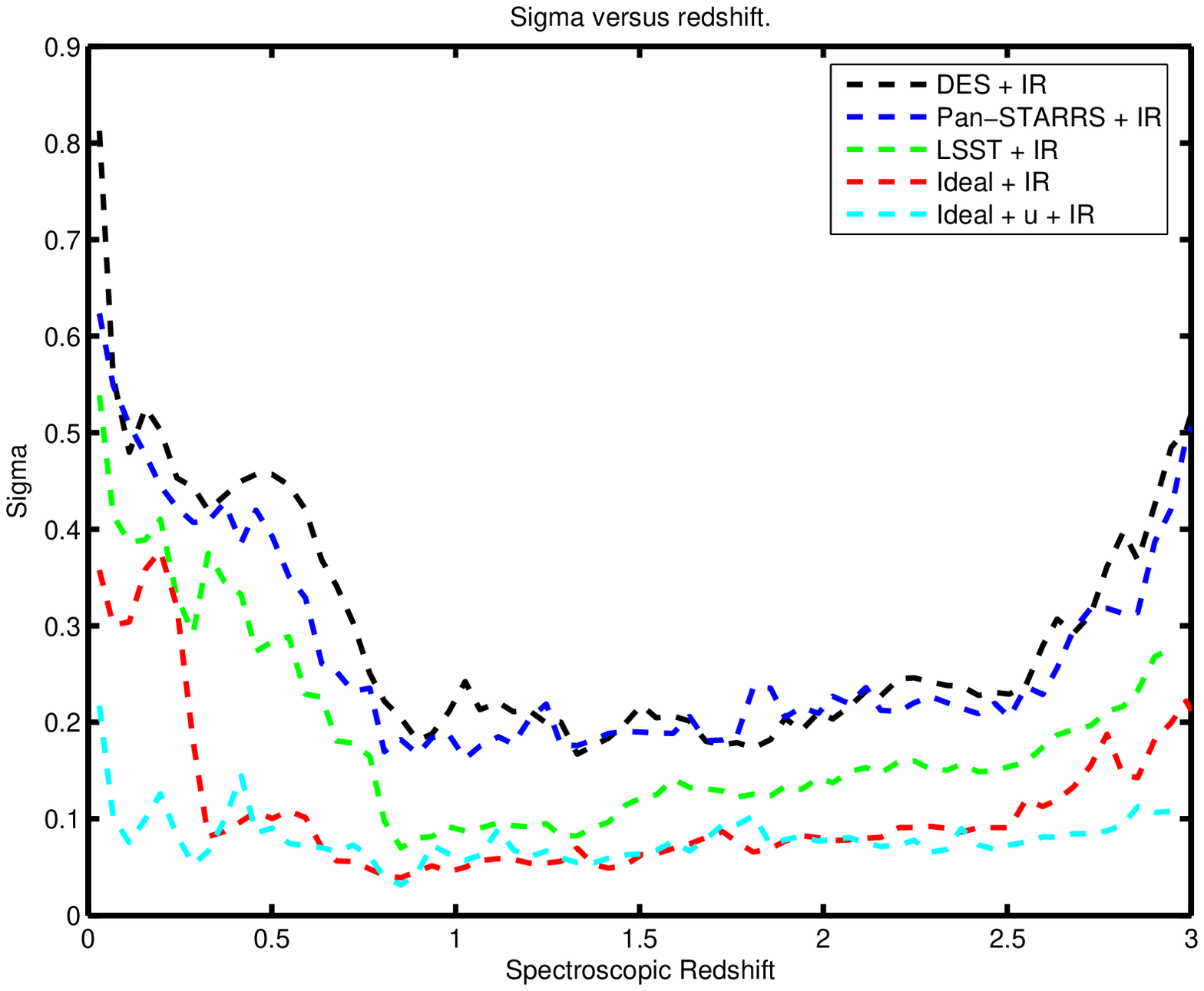}
\caption{The rms sigma ($\left<(z_{phot}-z_{spec})^2\right>$, bottom graphs)
and the bias ($\left<(z_{phot}-z_{spec})\right>$, top graphs)
as a function of redshift
for all the cases considered in this section (left without IR, right
including IR). We can see how the inclusion of
IR data improves significantly the data with optical exposure times
but helps less data with significant optical exposure times although there is
still improvement. We can also see the relative importance of bands by
comparing surveys with deeper y and z bands and surveys with u band data.
\label{fig:sig_bias}}
\end{center}
\end{figure*}

The remaining freedom left in a neural network analysis is the
architecture of the network. A simple architecture is easier to
minimise but may not provide the best fit to the data. On the other
hand a complicated architecture may remain stuck in a local minimum
of the cost function more easily and hence not provide the best
solution to the problem either. We do not attempt to optimise the
architectures for each scenario as we consider that this work does
not have as an aim to judge the performance of different photometric
redshift techniques. A network with architecture N:2N:2N:1 (i.e.
which has N inputs, two hidden layers with 2N nodes each and only
one output estimating the redshift, and where only adjacent layers
are interconnected) has been shown to work well on photometric data
\citep{2004PASP..116..345C} where N is the number of different
photometric bands available. Therefore we choose this architecture
in all our scenarios. For details about the architecture see
\citet{2004PASP..116..345C} and references therein.

\subsection{Impact of the photometry depth and bands}
\label{sec:impact_bands}

In this section we compare the photometric redshift quality we get
for different choices of survey depths and different choices of
filters for the simulations we have presented in Sec.\ref{sec::cat}.
The aim of this section is mainly to assess the impact of the u
band, IR bands and the redder optical bands (z and y) on the output
of a photometric redshift code.

\begin{table}
  \begin{center}
    \begin{tabular}{|l|c|c|c|c|}
      \hline
      Survey -- $\sigma$ & $z=[0,0.5]$  & $[0,1.5]$  & $[1.5,3]$ & $[0,3]$ \\
      \hline
      $RIZ$           & 0.874 & 0.643 & 0.728 & 0.668 \\
      $Des$           & 0.803 & 0.545 & 0.636 & 0.572 \\
      $Des + IR$      & 0.463 & 0.330 & 0.238 & 0.307 \\
      $Pan$           & 0.796 & 0.515 & 0.635 & 0.552 \\
      $Pan + IR$      & 0.428 & 0.289 & 0.233 & 0.274 \\
      $LSST$          & 0.663 & 0.392 & 0.429 & 0.403 \\
      $LSST + IR$     & 0.342 & 0.211 & 0.155 & 0.197 \\
      $Ideal$         & 0.517 & 0.296 & 0.310 & 0.300 \\
      $Ideal + IR$    & 0.213 & 0.119 & 0.097 & 0.113 \\
      $Ideal + u(25)$ & 0.375 & 0.218 & 0.217 & 0.217 \\
      $Ideal + u$     & 0.290 & 0.170 & 0.160 & 0.167 \\
      $Ideal + u + IR$& 0.101 & 0.074 & 0.080 & 0.076 \\
      \hline
      Survey -- $\sigma_{68}$ & $z=[0,0.5]$  & $[0,1.5]$  & $[1.5,3]$ & $[0,3]$ \\
      \hline
      $RIZ$           & 1.029 & 0.667 & 0.758 & 0.691 \\
      $Des$           & 0.748 & 0.346 & 0.559 & 0.412 \\
      $Des +IR$       & 0.238 & 0.147 & 0.175 & 0.155 \\
      $Pan$           & 0.728 & 0.327 & 0.548 & 0.391 \\
      $Pan + IR$      & 0.216 & 0.128 & 0.166 & 0.139 \\
      $LSST$          & 0.346 & 0.167 & 0.276 & 0.196 \\
      $LSST + IR$     & 0.122 & 0.085 & 0.119 & 0.094 \\
      $Ideal$         & 0.233 & 0.122 & 0.173 & 0.136 \\
      $Ideal + IR$    & 0.068 & 0.047 & 0.075 & 0.054 \\
      $Ideal + u(25)$ & 0.094 & 0.073 & 0.140 & 0.089 \\
      $Ideal + u$     & 0.057 & 0.052 & 0.122 & 0.067 \\
      $Ideal + u + IR$& 0.039 & 0.036 & 0.063 & 0.043 \\
      \hline
    \end{tabular}    \vspace{2mm}
  \end{center}
  \caption{The $\sigma_{68}$ (defined by the interval
in which 68 per cent of the galaxies that have smallest
$z_{phot}-z_{spec}$ lie within) and $\sigma$ (defined by
$\left<(z_{phot}-z_{spec})^2\right>$) for different choices of
depths and filters. A comparison of these two numbers assesses how
many outliers there are in the sample and whether the distribution
of photometric redshifts for a given spectroscopic redshift interval
is Gaussian or has broad tails. We note that in order to obtain
reliable photo-z for galaxies with an RIZ depth of 25, shallower
surveys such as DES or Pan-STARRS are not well matched to DUNE,
hence the scatter is large. We note that shallower surveys would not
use such faint galaxies for lensing. Deeper optical surveys are
necessary to reproduce good photo-z on a galaxy-by-galaxy basis,
however IR data considerably improves even the shallower surveys. We
also include one line based on photometric redshifts from RIZ band
only, as a baseline for comparison to other surveys and one line
assuming an ideal survey with a shallower u band down to 25 so
illustrate the increment accuracy that the u band gives from 24 to
26. \label{tab::table_sigma_2}}
\end{table}

\begin{figure*}
\begin{center}
\begin{minipage}[c]{1.00\textwidth}
\centering
\includegraphics[width=8.7cm,angle=0]{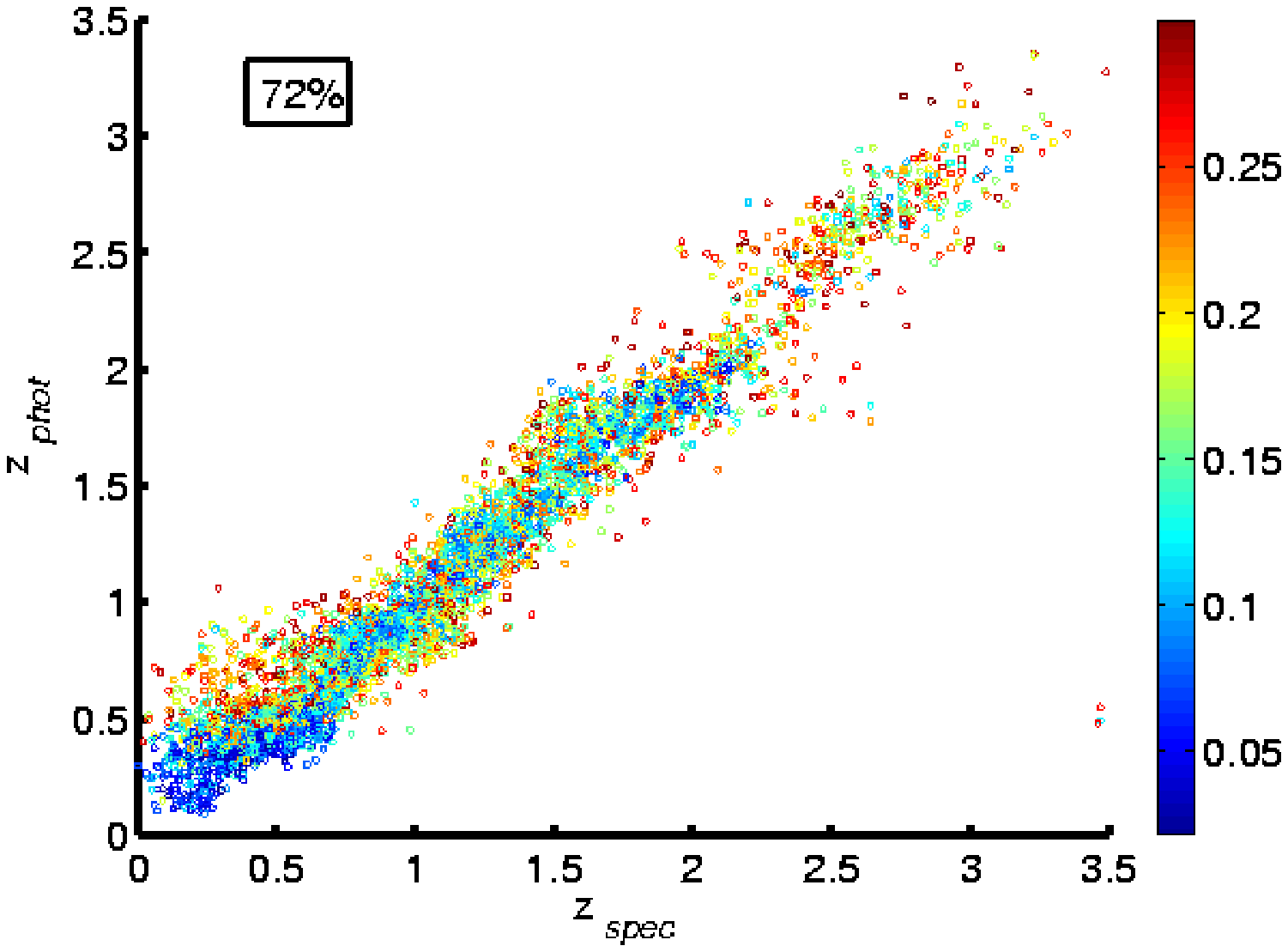}
\includegraphics[width=8.7cm,angle=0]{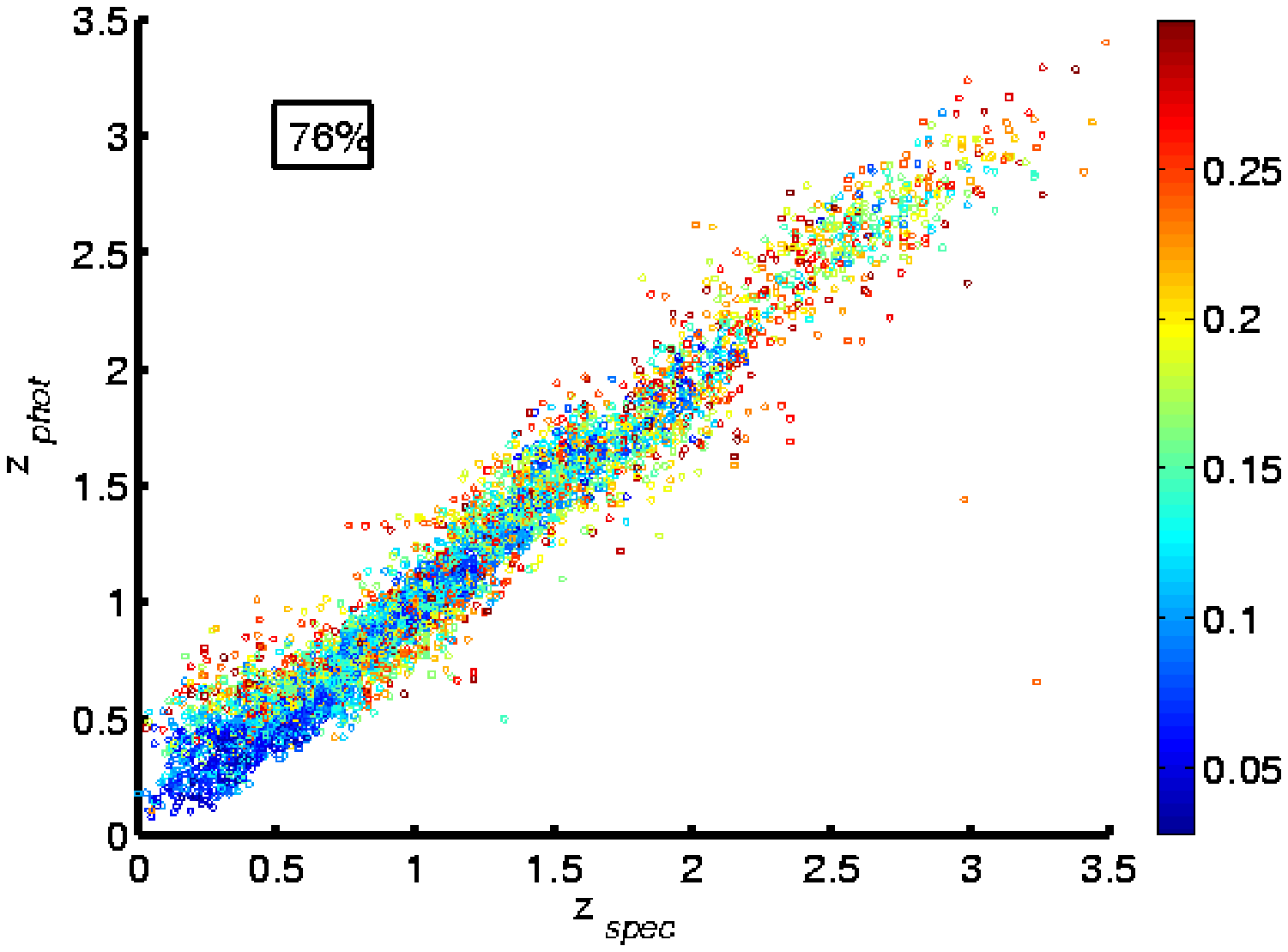}
\includegraphics[width=8.7cm,angle=0]{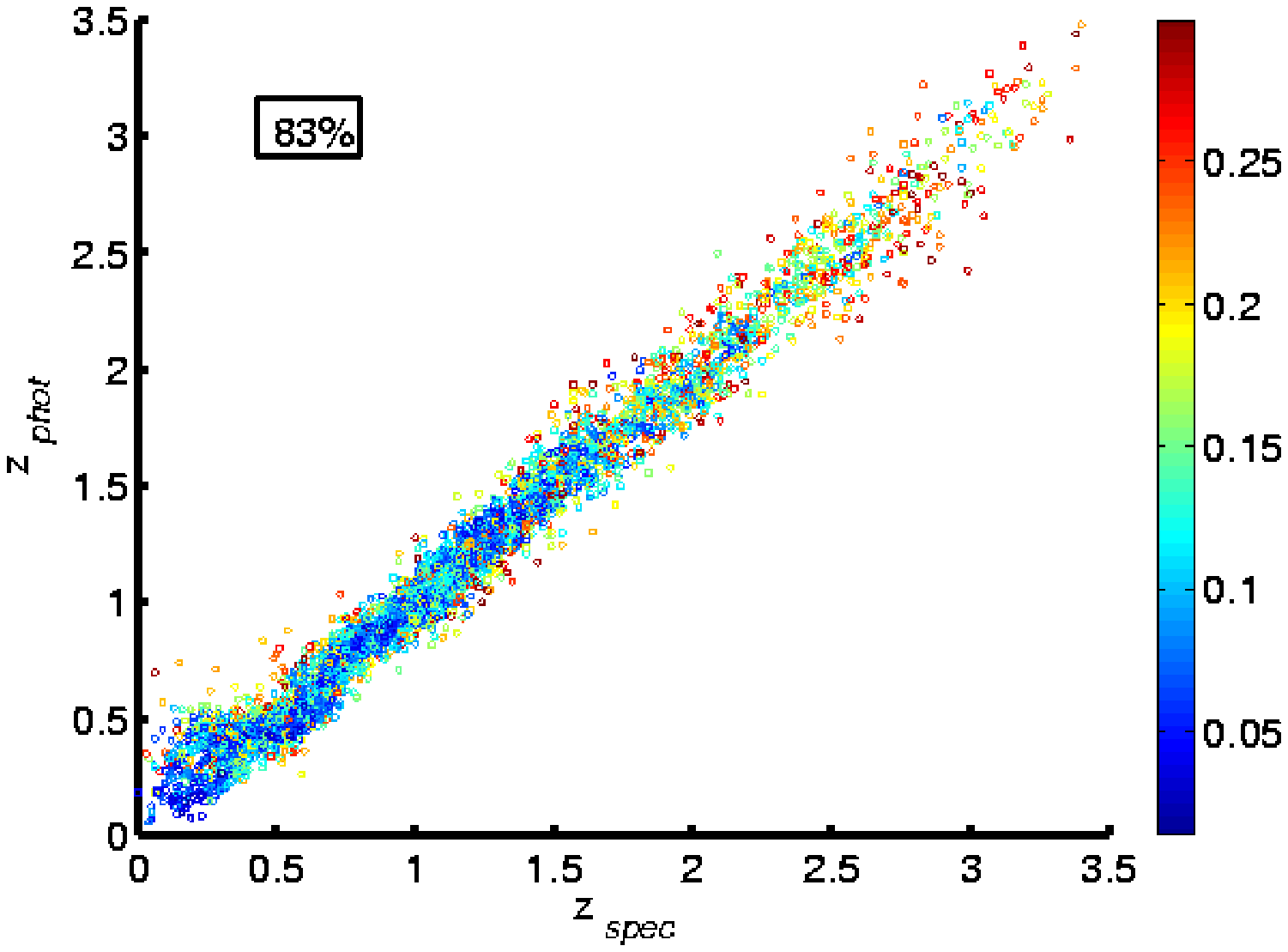}
\end{minipage}
\begin{minipage}[c]{1.00\textwidth}
\centering
\includegraphics[width=8.7cm,angle=0]{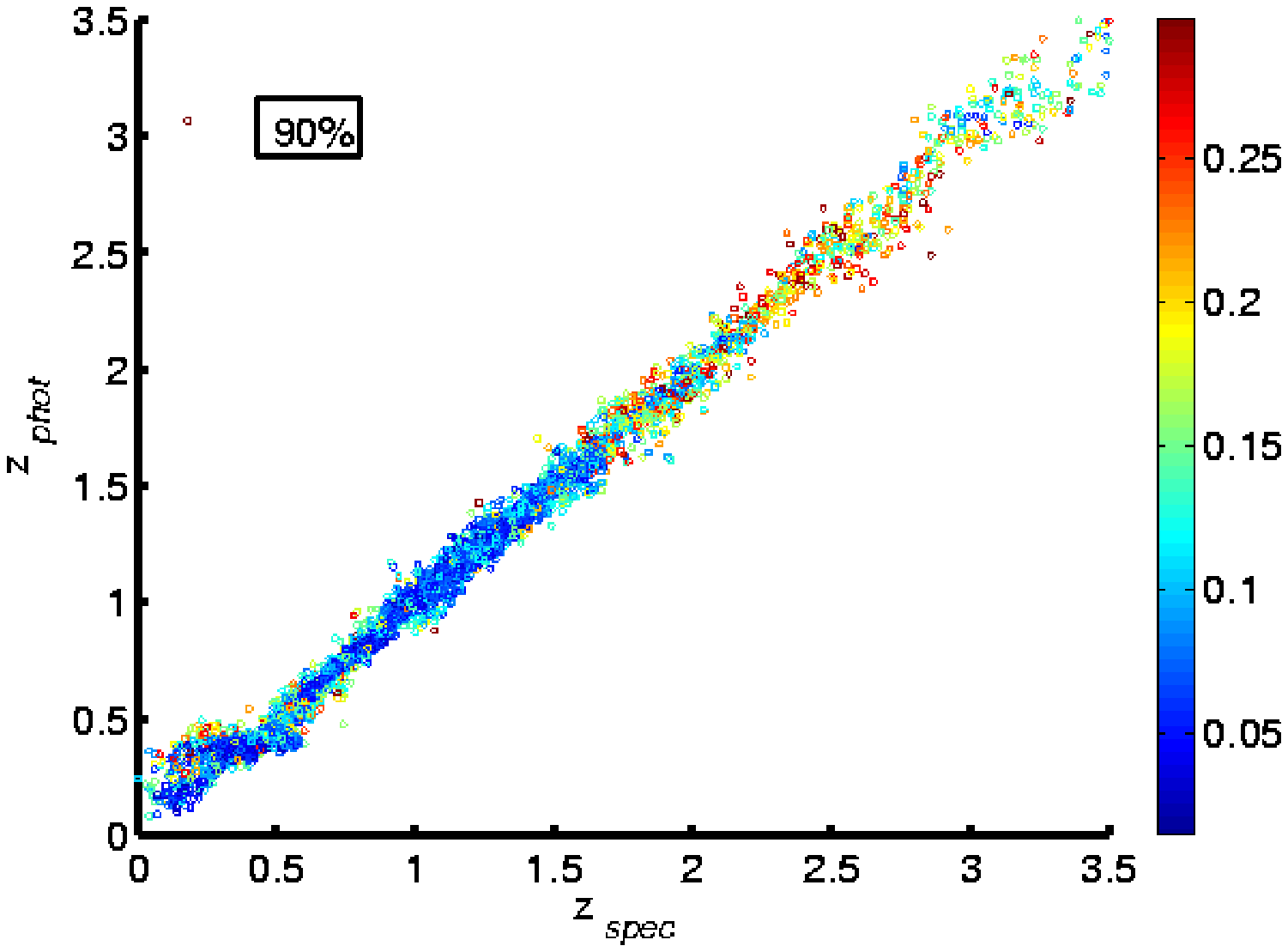}
\includegraphics[width=8.7cm,angle=0]{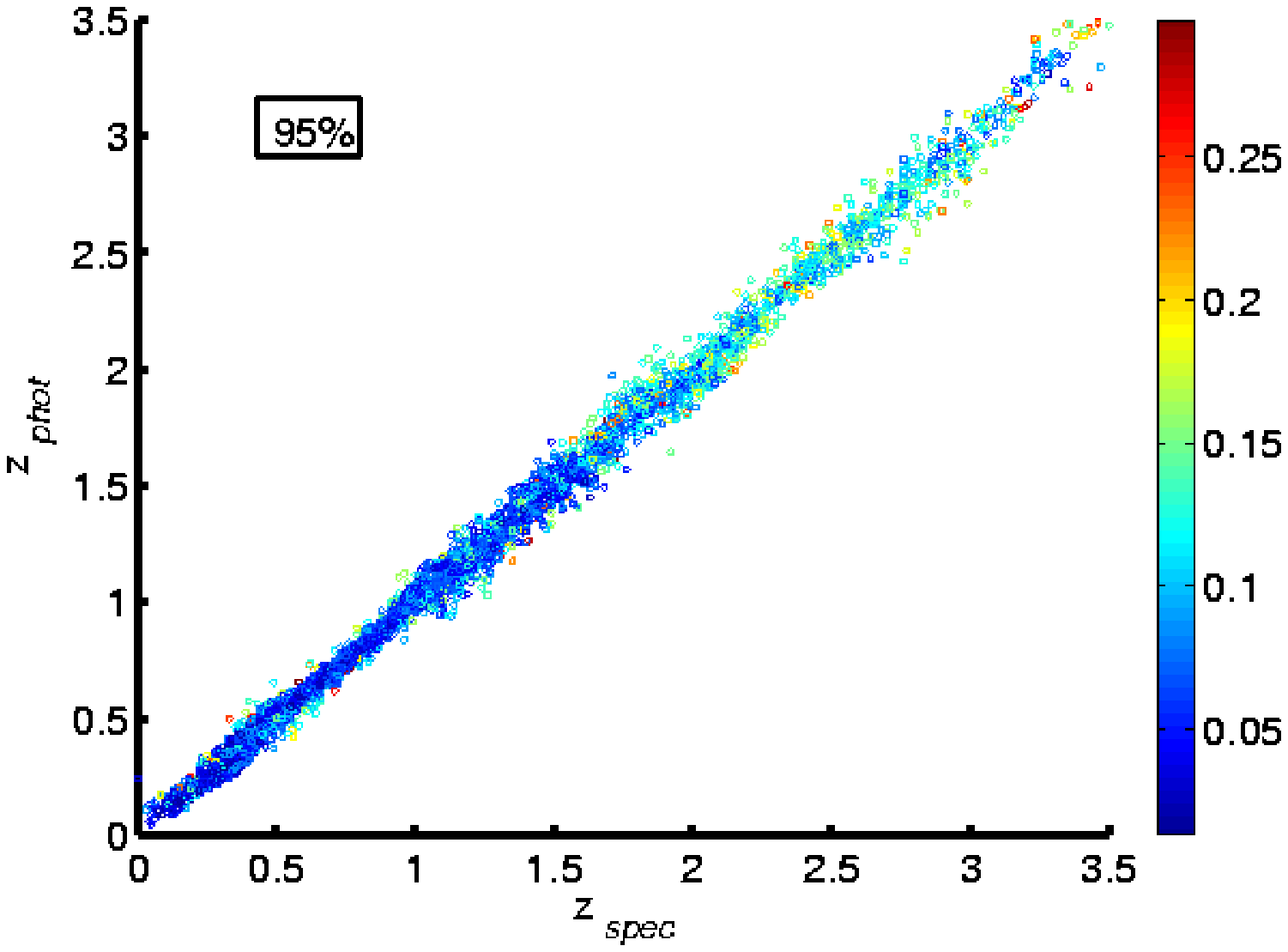}
\end{minipage}
\caption{Cleaned catalogues for the 5 surveys
considered before. All surveys contain IR information. As  explained
in Sec.\ref{sec:clean}, we have removed all photometric redshifts with an error
estimate larger than 0.3. After a neural network is trained it can
assess the error on the photometric redshift looking at the error
in in the photometry, this value is used for the cut presented here.
We can see from all the scatter plots
that although the error estimate from the neural networks is not the
best available it is able to remove the correct galaxies and provide
correct photo-z for some of the sample. After the cut, only 72\% of
the galaxies were left on the DES+ IR catalogue; 76\% had good photo-z
on the Pan-STARRS + IR catalogue; 82\% of the sample on the LSST + IR
catalogue; 90\% and 95\% of the sample on the Ideal + IR and Ideal + u + IR
catalogues respectively. As we can see from Fig.\ref{fig:error} a clipping of
0.3 is indeed conservative, we can clean the different
catalogues at a higher error estimate but we make a comparison
between catalogues here. We stress that this is not a sigma clipping, we do not
use information about the spectroscopic redshift of the sample to make
this cleaning procedure, we only use the photometry. The colour bar indicates
the error estimate given by the neural network code.
\label{fig:clean}}
\end{center}
\end{figure*}

\begin{figure*}
\begin{center}
\begin{minipage}[c]{1.00\textwidth}
\centering
\includegraphics[width=8.7cm,angle=0]{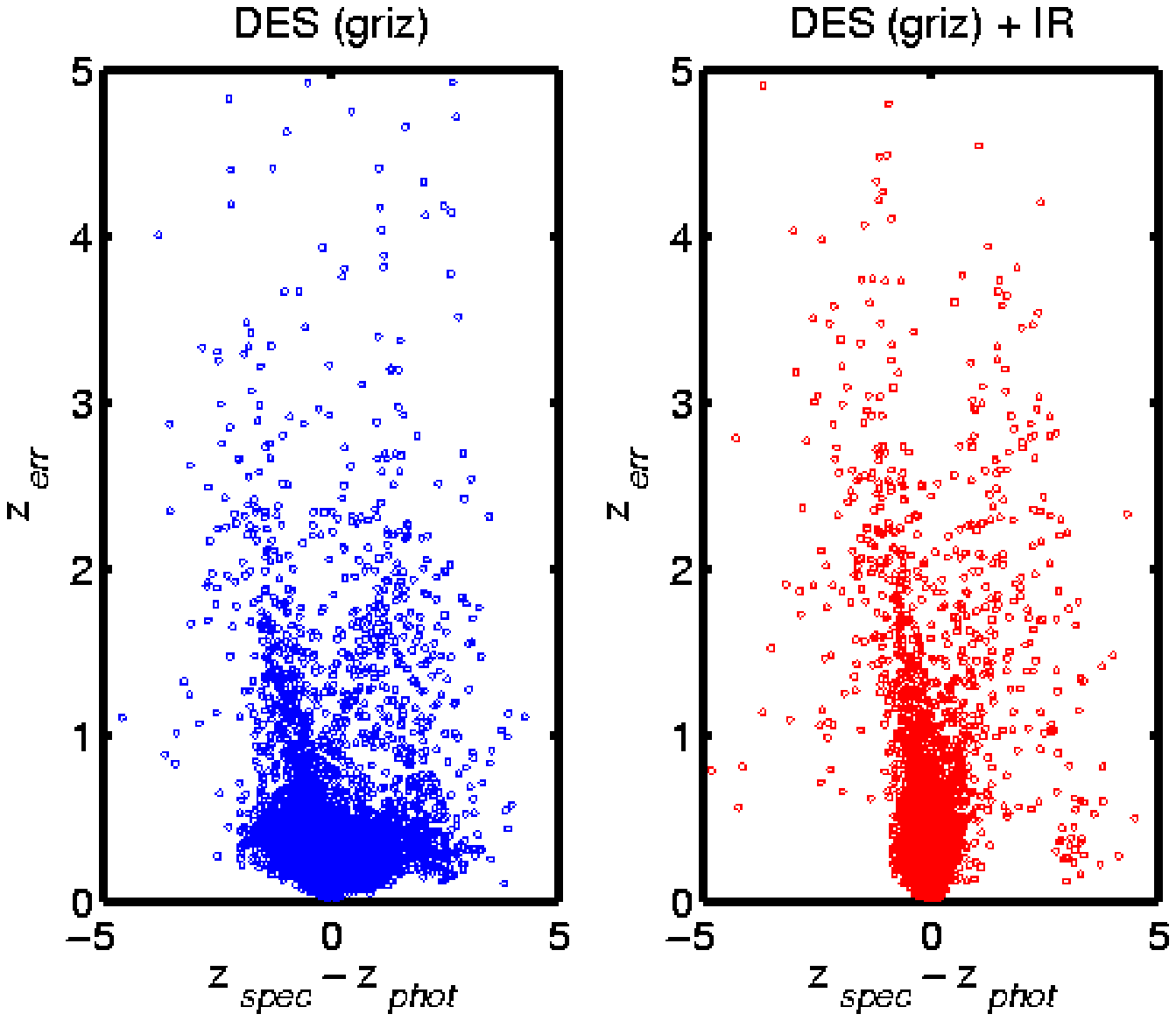}
\includegraphics[width=8.7cm,angle=0]{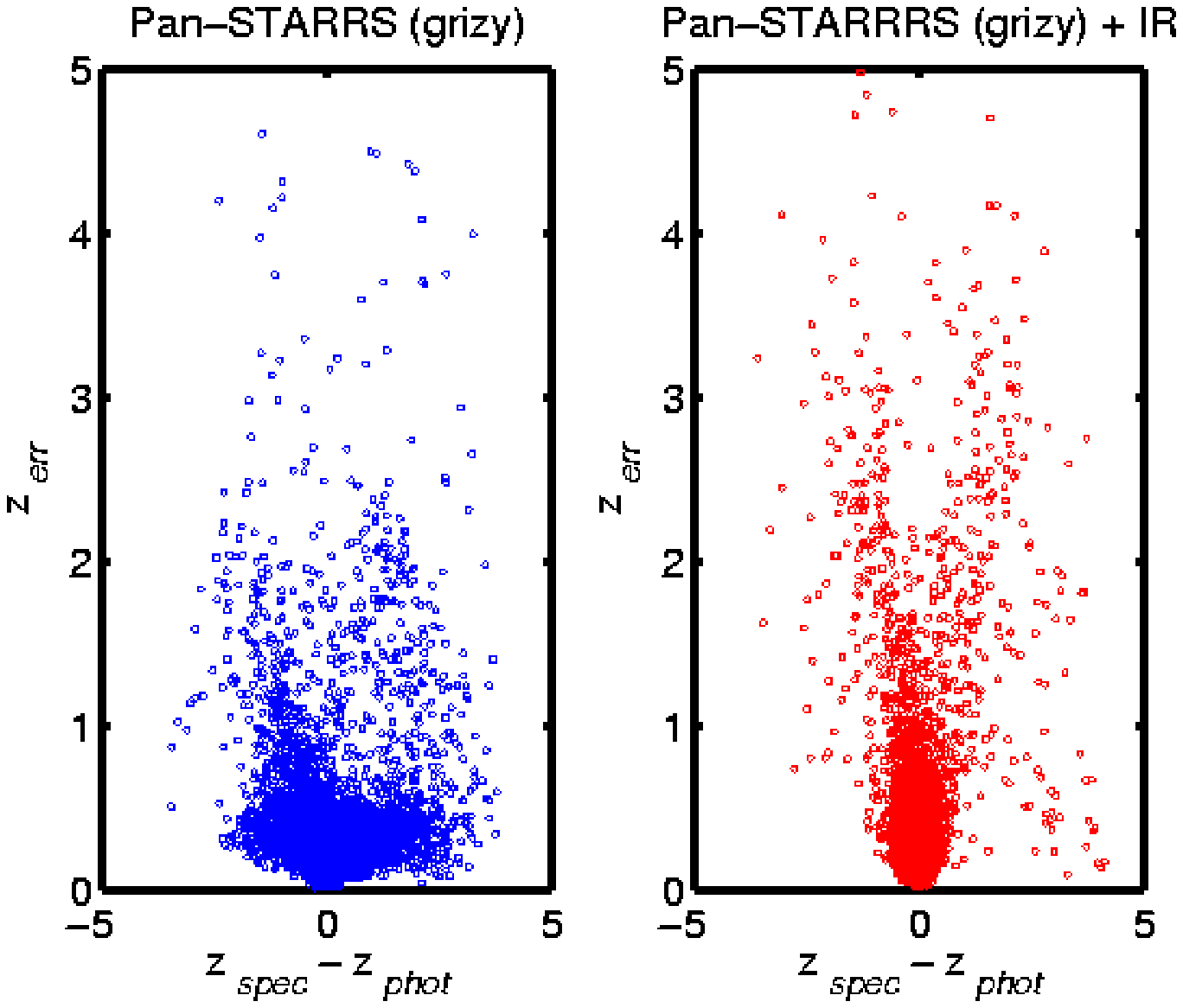}
\includegraphics[width=8.7cm,angle=0]{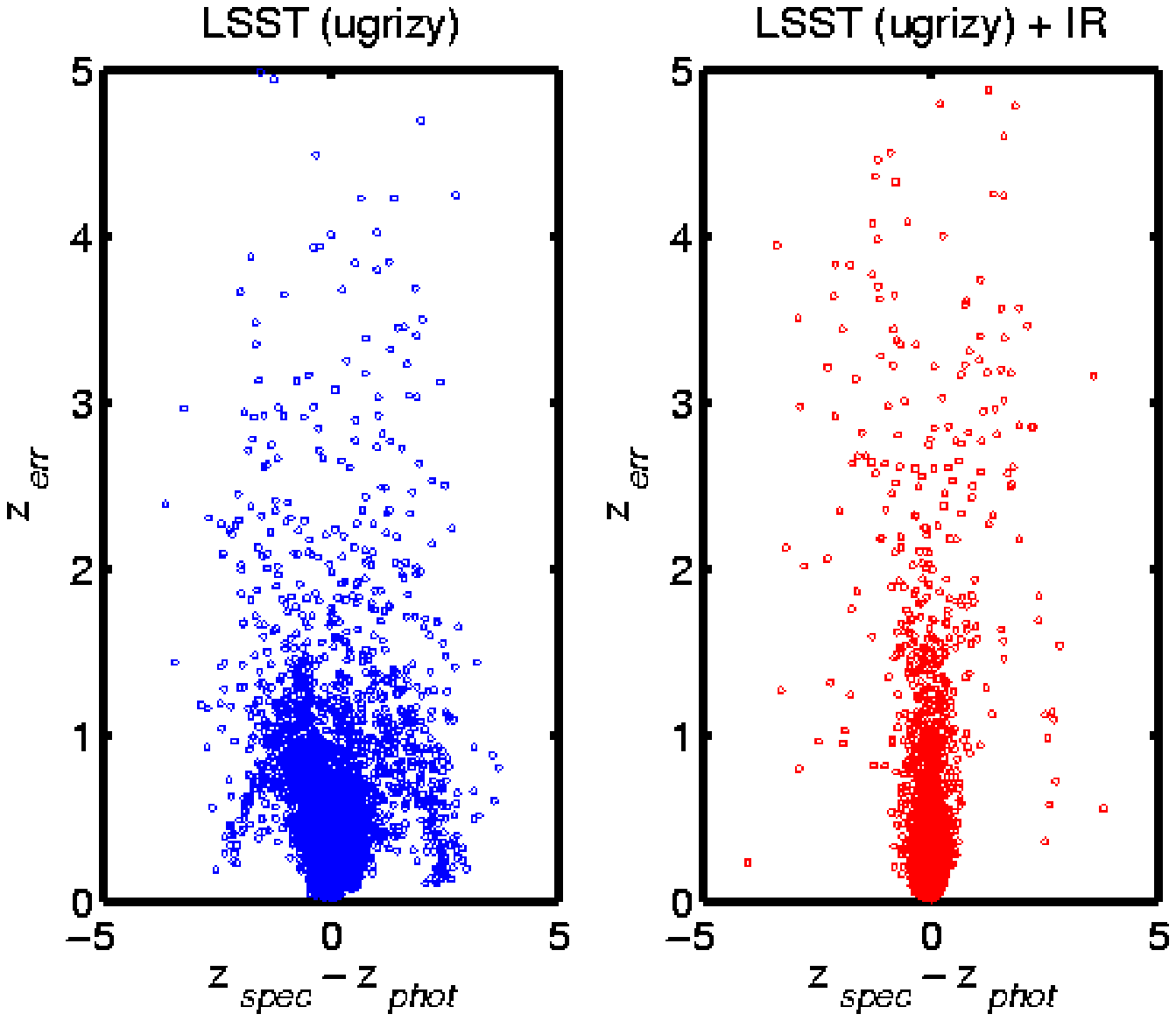}
\end{minipage}
\begin{minipage}[c]{1.00\textwidth}
\centering
\includegraphics[width=8.7cm,angle=0]{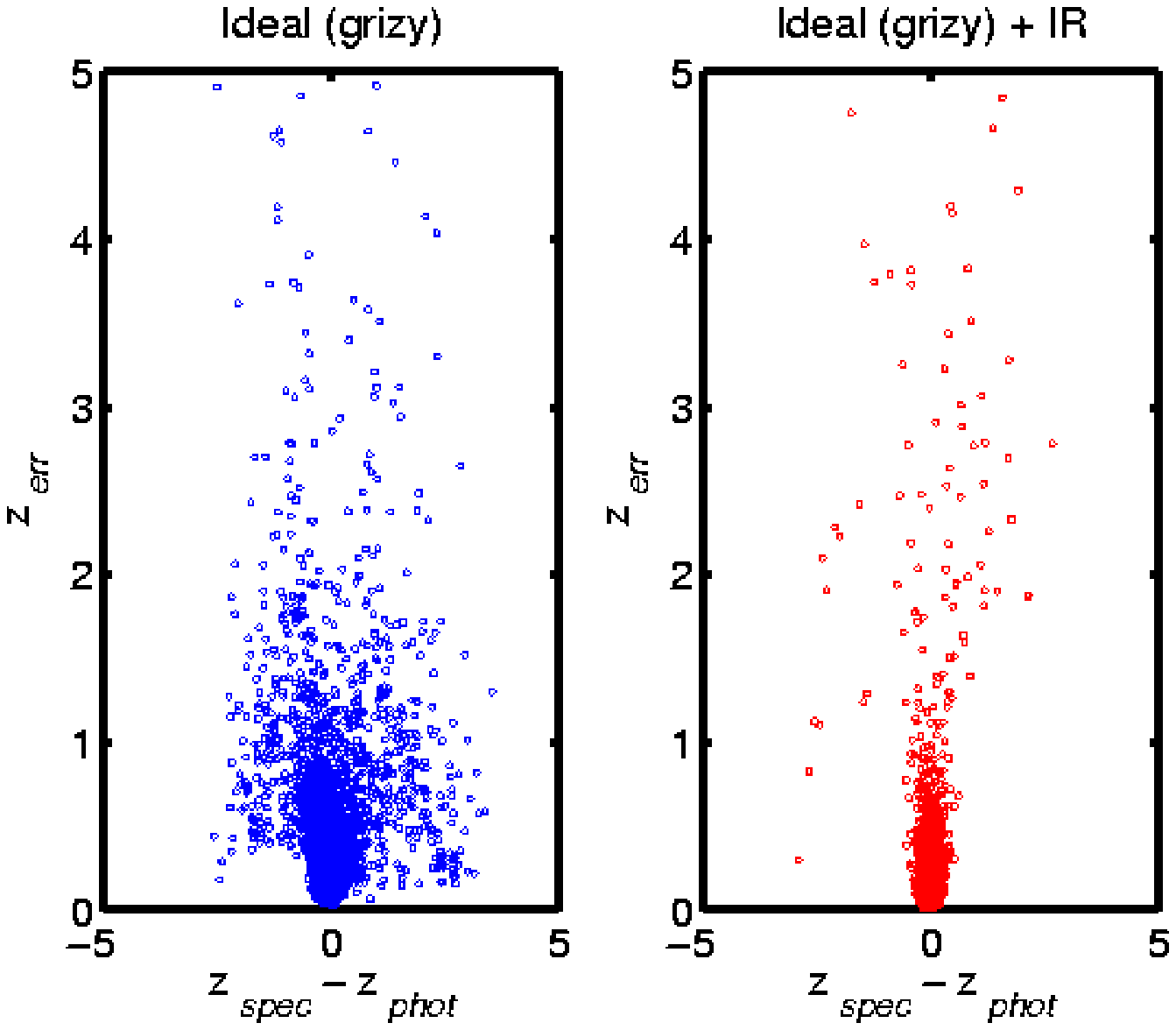}
\includegraphics[width=8.7cm,angle=0]{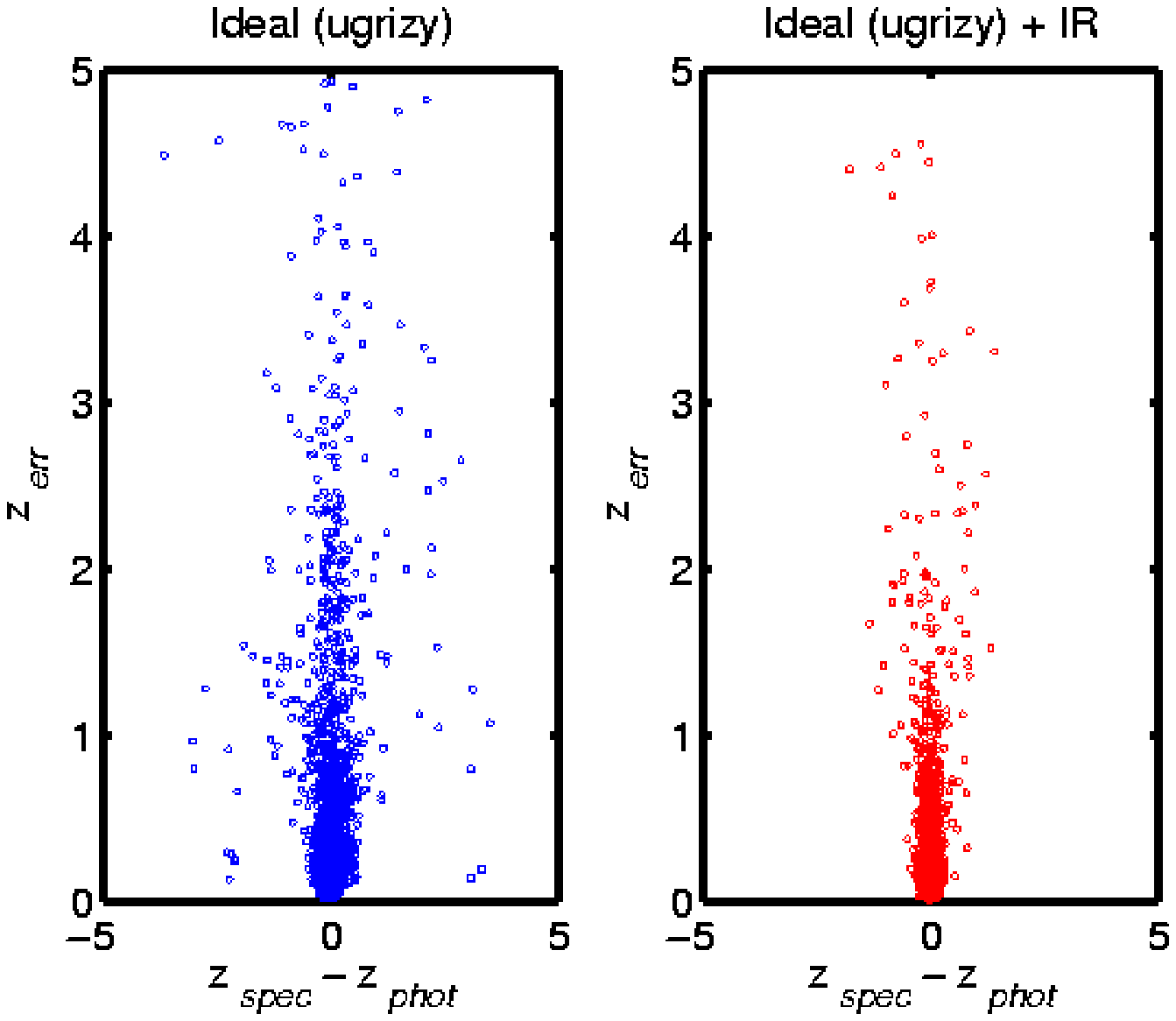}
\end{minipage}
\caption{The relation between $z_{phot}-z_{spec}$ as a function of the
error predicted by the neural network $z_{err}$ based on the errors
from photometry. We can see that the scatter points are not distributed
in a Gaussian way around the centre for the shallower surveys and this becomes
more apparent in the deeper surveys; hence the error estimate becomes more
reliable the deeper the survey is. Furthermore, the error estimate is
more reliable when IR data or u band data is obtained.
We also note that in the specific
case of neural network error estimates, with IR data the error is somewhat
overestimated. These scatter plots show that in the case of DES + IR a
cut at 0.3 is appropriate to remove most outliers.\vspace{10ex}
\label{fig:error}}
\end{center}
\end{figure*}

\begin{figure*}
\begin{center}
\includegraphics[width=11.5cm,angle=0]{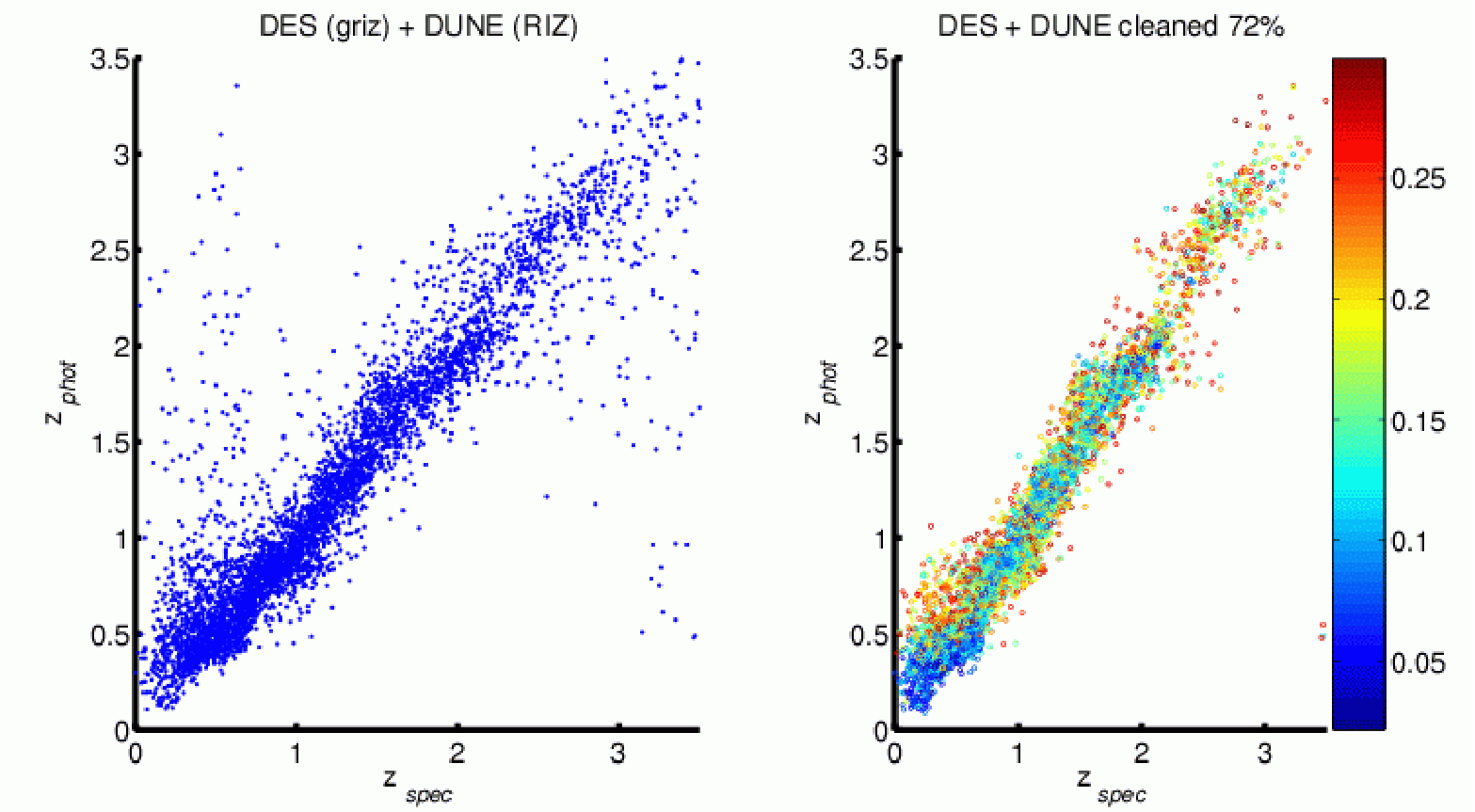}
\includegraphics[width=11.5cm,angle=0]{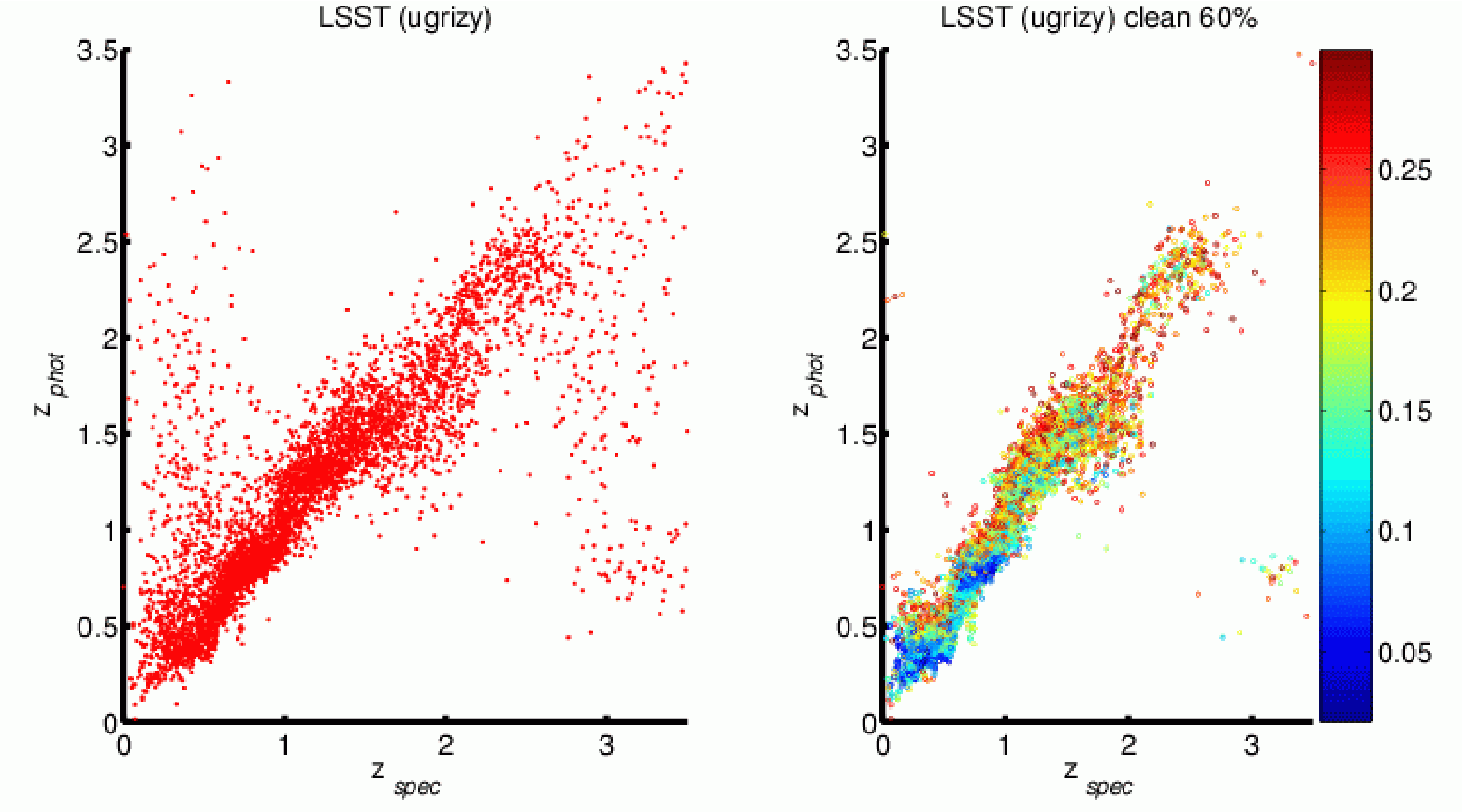}
\includegraphics[width=11.5cm,angle=0]{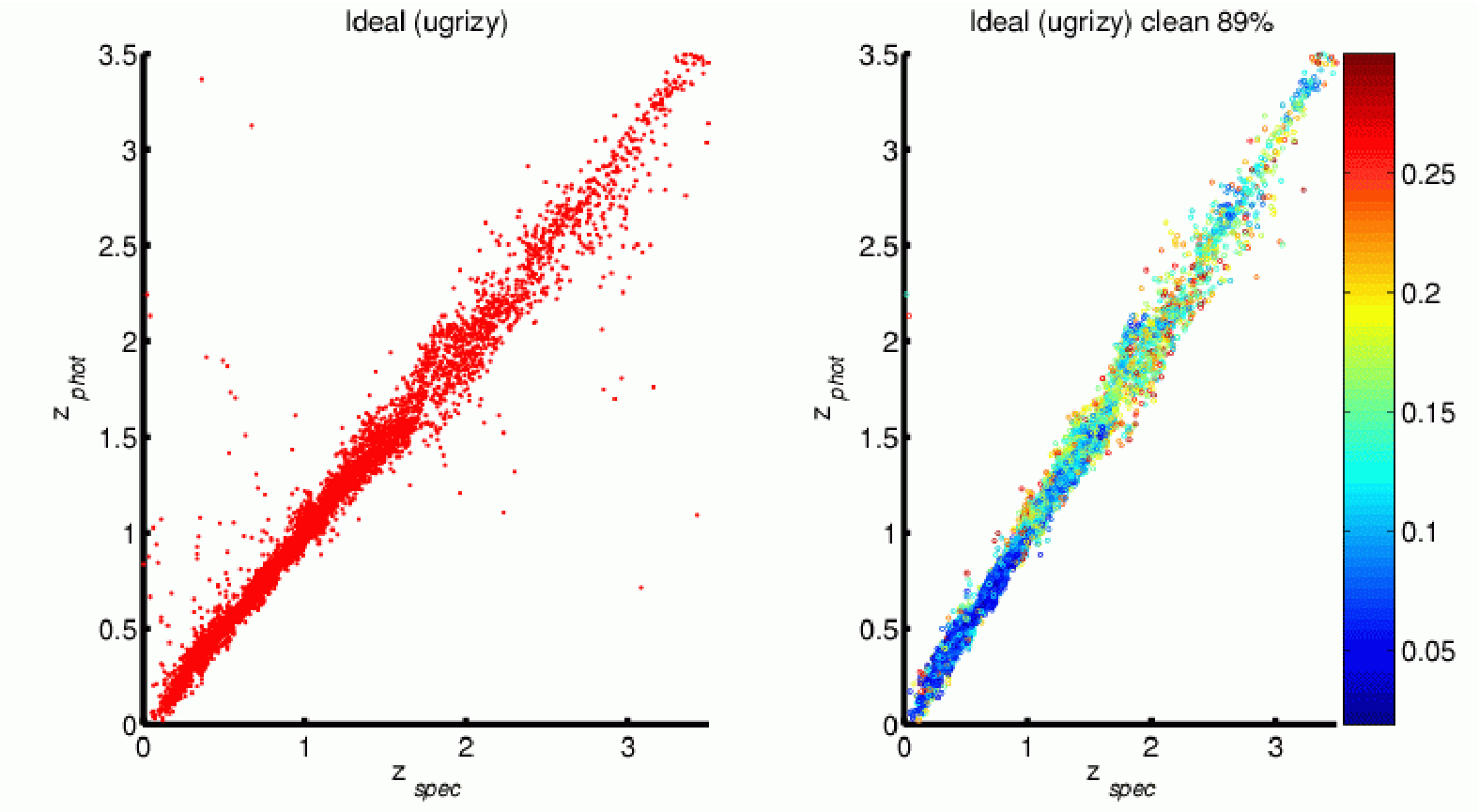}
\caption{We present here scatter plots for some configurations
before and after the cleaning. The top panels are in the case of DES
photometry plus DUNE photometry, the middle panel is for LSST depth
photometry without IR and the bottom panels are for the case of an
ideal optical survey with very deep u band photometry. We can see
that the cleaning is very efficient and removes outlier galaxies
without removing good photo-z where as the same is not entirely true
for the case of LSST photometry alone. For instance at some
redshifts (around 3) some galaxies with good photo-z are removed
where some outliers remain after the cleaning. The effect is much
worse for poorer optical photometry than LSST for depths of 25 in
the RIZ band. However we can see from the bottom panel that deep u
band photometry can cure this. The conclusion is that cleaning can
be an effective method to remove outliers but it is only as
effective as the baseline of wavelengths available. Without deep IR
or u band data this can lead to bad photo-z error determinations.
\label{fig:error_compare}}
\end{center}
\end{figure*}

 We have chosen five optical baseline surveys. One which has a depth
equivalent to what the Dark Energy Survey (DES) will be able to
provide, roughly going down to a magnitude limit of 24 in the four
bands (griz); another is an equivalent to what the collection of
four Pan-STARRS telescopes will provide, potentially obtaining an
order of magnitude increase in depth in the five bands (grizy); the
third is an estimate of what an optical survey with a Large Synaptic
Survey Telescope (LSST) would achieve, obtaining another order of
magnitude increase in depth in 6 bands (ugrizy). We have also
considered two ideal optical surveys, one without u band photometry
but with a very deep exposure in the z and y bands; another which is
very similar to the first but with very deep u band imaging. The
depths of theses surveys are outlined in Table.\ref{tab::depths} and
are chosen to be roughly consistent with the limits which the Dark
Energy Survey\footnote{http://decam.fnal.gov} (which will survey the
sky with the 4m Blanco telescope), Pan-STARRS
4\footnote{http://pan-starrs.ifa.hawaii.edu} (which will be a
collection of four 2m telescopes) and
LSST\footnote{http://www.lsst.org} (a project to survey the sky
every night with a large 8m telescope) will be able to attain
although the conclusions of this study are not dependent on the
particulars of these projects.

We measure photometric redshifts with these baseline surveys
both including and
excluding the additional information that a space based mission
would add with infrared detectors. We plot our findings in
Fig.\ref{fig:zin_zout_IR} and Fig.\ref{fig:zin_zout_IR_dens}
and show the scatter in redshift intervals
in Table.\ref{tab::table_sigma_2} and Fig.\ref{fig:sig_bias}. The sigma
is the r.m.s. photometric redshift error around the mean, and
$\sigma_{68}$, the interval in which 68 per cent of the galaxies have
the smallest value of $z_{spec}-z_{phot}$.
We can see from the blue samples of
Fig.\ref{fig:zin_zout_IR} how the increasing depth of the optical
survey is significant in obtaining photometric redshifts. We find that
to obtain reliable photometric redshifts for a sample with an RIZ
magnitude below 25 we are still improving significantly the
quality of the photometric redshifts if the overall photometry is
as deep as 26 or larger. We note that in order to obtain reliable photo-z
for galaxies with an RIZ depth of 25 shallower surveys such as
DES or Pan-STARRS are not well matched to DUNE, deeper surveys
are necessary to reproduce good photo-z on a galaxy-by-galaxy basis,
however IR data considerably improves even the shallower surveys.

We can also see, by analysing these scatter plots, the relative
importance of different bands. We can assess, for instance, how much
the deeper exposure times in the z and y bands would help in
producing reliable photometric redshifts compared to near infrared
data; if we compare the LSST + IR case with the ideal case without u
band we can see that out to redshift $z \sim 1.6 - 1.7$ the deeper z
and y bands help and the improvement due to IR data is not as large
as when the depths in y and z are shallower, however, for
photometric redshifts of galaxies above a redshift of $z \sim 2.5$
are only improved by the inclusion of IR data. We can understand
this behaviour of the error in the photometric redshift estimate
with the following argument: most of the information in photometric
redshifts come from the 4000 \AA~ break in galaxy spectra, hence the
best photometric redshift estimates are expected at the redshifts
where the 4000 \AA~ break falls between bands with deep exposures.
We can see that optical surveys have the best photometric redshift
estimates around a redshift of $z \sim 0.7$ which corresponds to the
4000 \AA~ break falling in the i band, hence having measurements of
the spectrum on both sides of the break. Similarly, galaxies at a
higher redshift are helped by the measurement of the break
redshifted into the IR bands. This argument can also be applied to
the Lyman break which occurs at 912 \AA. This break is redshifted to
the blue part of the spectrum for galaxies around redshift $z \sim 2
- 2.5$, hence u band data also helps photometric redshifts for
galaxies in this region of the spectrum.

\begin{figure}
\begin{center}
\includegraphics[width=8.5cm,angle=0]{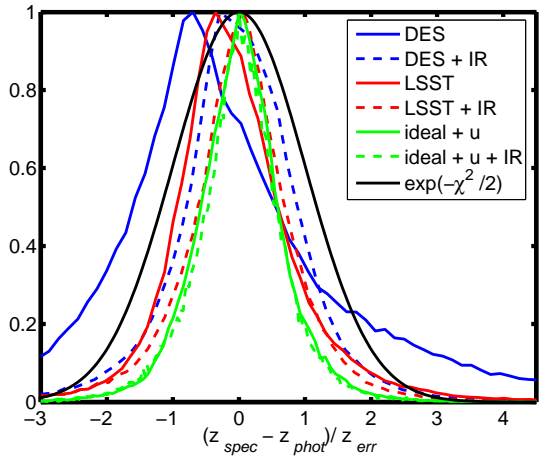}
\caption{Fit to the unnormalised histograms of the quantity
$(z_{spec}-z_{phot})/z_{err}$. If the error estimator is reliable,
the quantity above being one corresponds to the one sigma error
limit. Therefore 68\% of the points should have
$(z_{spec}-z_{phot})/z_{err}$ smaller
than one in absolute value
and centred around zero. We plot this fit for six cases
described in Sec.\ref{sec:impact_bands} with and without IR data. We can see
that for a cut in RIZ at magnitude 25, the shallower surveys produce data
which leads to a
biased error estimates with neural networks. All cases with IR
data have a much better error estimates and therefore cleaning can be
done more efficiently. We can see that the error estimates are less biased
with the LSST case at this depth and are not biased at all if we have
deeper u, z and y data. We also plot the curve $exp(-x^2/2)$ for
comparison. If the photo-z error estimate is reliable, then the distributions
shown here should be close to Gaussian. We can see that the DES mock without IR
(blue curve) does not yield very reliable photo-z error estimates whereas
the other configurations, including the DES mock with IR, produce more reliable
results.
\label{fig:hist_err}}
\end{center}
\end{figure}

\begin{figure*}
\begin{center}
\includegraphics[width=19.0cm,angle=0]{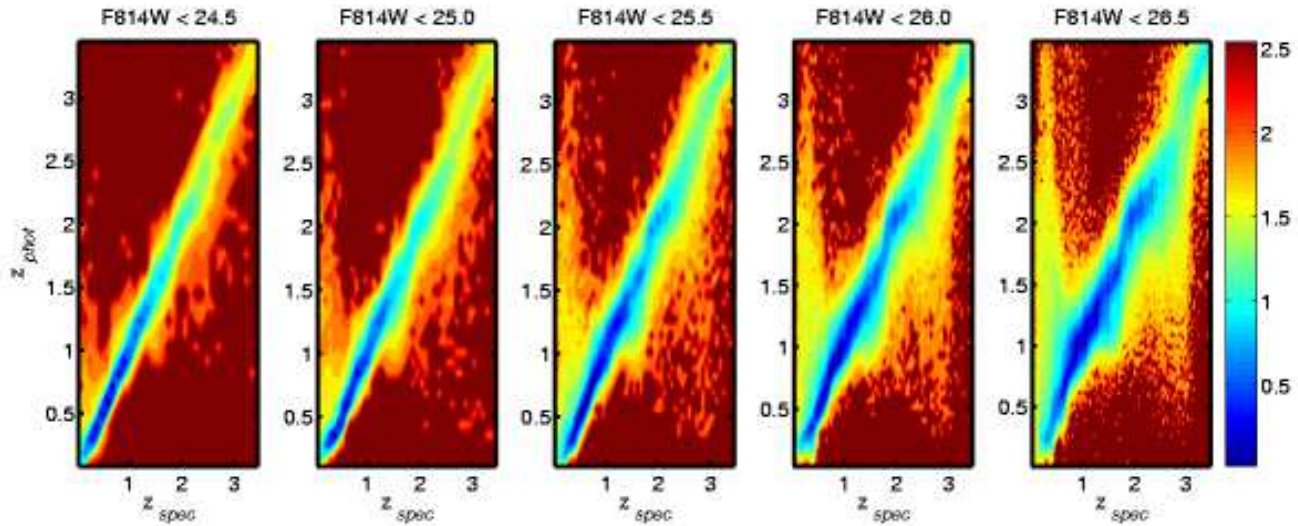}
\caption{Density plots in the $z_{phot} - z_{spec}$ plane for
increasing depth of a catalogue. We have taken mock catalogues
derived from the Cosmos survey and we made cuts in the filter F814W
at several magnitude from 24.5 to 26.5 in steps of 0.5. We have
assumed complete training sets with the same number density of
training galaxies per colour volume for every case shown here. We
can see that the bright galaxies with high signal to noise have well
constrained photometric redshifts whereas the noisy galaxies have
almost unconstrained photometric redshifts. The plots are colour
coded and the scale is exponential; a colour difference
corresponding to one is equivalent to the density being decreased by
a factor of $e$. \label{fig:cosmos_dens}}
\end{center}
\end{figure*}

The addition of u band data can significantly help the photometric
redshift determination of certain galaxies. This information is
helpful in removing catastrophic outliers that have a small
spectroscopic redshifts and which have relatively featureless SEDs.
We can see, comparing LSST with DES or Pan-STARRS, that there are
indeed fewer outliers if we include u band imaging to our mock
survey. However, comparing the LSST simulated results with the
idealised survey including much deeper u band imaging we see that
the u band depth has to be larger then 24.0 if we want all the
catastrophic outliers to be removed for a RIZ survey below 25.0; at
a magnitude limit for the u band of 26.0 almost all the catastrophic
outliers have been removed, we also provide numbers for a similar
survey with u band as deep as 25 to illustrate the increment given
by the u band in the range 24.0 to 26.0. We note that these
integrations on the u band are extremely difficult verging on the
impossible even with the next generation surveys. We also note that
the inclusion of u band data has a similar effect to the inclusion
of IR data in the case of an idealised survey; the scatter including
IR data is still significantly improved (see
Table.\ref{tab::table_sigma_2} and Fig.\ref{fig:sig_bias}), but
there are not many outliers with optical data only.

We have also assessed in Table.\ref{tab::table_sigma_2}, the amount
of outliers we get from a given survey configuration. We have
calculated $\sigma$ and $\sigma_{68}$ for each simulation. Whereas
the standard deviation gives us the spread for the entire sample,
$\sigma_{68}$ does so only for 68\% of the galaxies which have the
best photometric redshfits. Hence $\sigma_{68}$ is insensitive to
outliers whereas $\sigma$ is very sensitive to them.

\begin{table}
  \begin{center}
    \begin{tabular}{|l|c|c|c|c|}
      \hline
      $Band$ & $\rm Time$  & $\rm Overhead$ & $\rm Mapping$ & $\rm Total$ \\
      \hline
      $u$ & 3.80 & 1.00 & 1 & 3.80 \\
      $g$ & 1.00 & 1.00 & 1 & 1.00 \\
      $r$ & 1.58 & 1.01 & 1 & 1.60 \\
      $i$ & 3.02 & 1.03 & 1 & 3.12 \\
      $z$ & 7.59 & 1.10 & 1 & 8.32 \\
      $Y$ & 20.9 & 1.02 & 5.09 & 106.9 \\
      $J$ & 19.1 & 1.02 & 5.09 & 97.5 \\
      $H$ & 25.1 & 1.45 & 5.09 & 182.7 \\
      $K$ & 27.5 & 1.45 & 5.09 & 200.3 \\
      \hline
    \end{tabular}    \vspace{2mm}
  \end{center}
  \caption{Time cost for obtaining data in different bands
based on MEGAPRIME and WIRCAM on the Canada-France-Hawaii telescope
(CFHT) relative to g band on MEGAPRIME. The time factor is the
increase in integration time required to reach the same sensitivity
as g band due to background and instrumental sensitivity. The
overhead factor is the increase in overhead compared to the g band
taking into account the readout time of the instrument and maximum
integration times due to the sky background. The mapping factor is
the increase in integration time required to map the same area due
to the fact that IR instruments have a smaller field of view (due to
the cost of the detectors). The Total is the relative expensive in
total telescope time to reach the same depth as g band from the
ground. This table illustrates how much time it is required to
obtain IR data from the ground compared to optical
data.\label{tab::table_times}}
\end{table}

We note here, for example with LSST, that the u band requires 19
times as much integration time relative to other bands (specifically
g) to reach the same depth.  However blue optimised systems such as
LBT or CFHT can require as little as 4 times as much integration
time in U to reach the same depth (depending on the exact filter
choices) (see Tab.\ref{tab::table_times} for similar factors for
other bands based on CFHT and WIRCAM).  For an optimised space based
system using UV sensitive CCDs this factor would drop to around 2.
Specifically in the case of IR detectors; they are ten times more
expensive (in term of cost) than CCDs to cover the same area. 
However in terms of integration time they reach the same depth in
about the same amount of time. From the ground IR detectors are much
less sensitive due to the higher backgrounds. 

\subsection{Cleaned catalogues}
\label{sec:clean}

When a neural network is trained, we obtain an estimate for the
error on each of the photometric redshifts predicted. This error is
obtained the following way. For every scenario, the inputs of a
neural network have an associated noise to them. We can assess the
variance that this noise would introduce into the output of the
network by changing the inputs according to the error. This will
lead the the following error estimate:


\begin{figure}
\begin{center}
\includegraphics[width=8.5cm,angle=0]{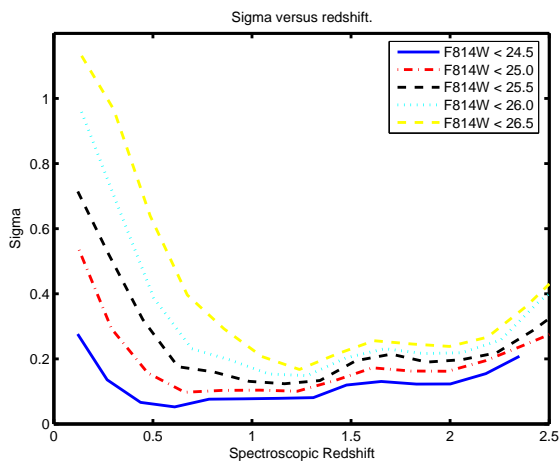}
\caption{The rms scatter of ($z_{phot}-z_{spec}$)
as a function of redshift for samples
with different magnitude cuts. Same data as in Fig.\ref{fig:cosmos_dens}.
As we can see
brighter samples have less galaxies but also have much better
photometric redshifts with many less outliers and smaller scatter.
\label{fig:cosmos_sig}}
\end{center}
\end{figure}

\begin{equation}
\sigma^2_z = \sum_i \left(\frac{\partial z}{\partial m_i}\right)^2
\sigma^2_{m_i}
\end{equation}

\noindent where the sum over i is a sum over all the network inputs.

In order to obtain the quantity $\partial z/\partial m_i$ we can use
an algorithm using the activation function for the weights of the
network described in \citep{Bishop}. This algorithm is fully
incorporated in the ANNz package \citep{2004PASP..116..345C} which
we use here.

We have plotted in Fig.\ref{fig:clean} the IR catalogues cleaned
conservatively with a photometric error estimate of 0.3. We have
chosen this threshold as it removes a vast majority of the outliers
in the shallower catalogue (DES + IR). In this case we can see that
around 30\% of the sample has been removed from the data however the
quality of the photometric redshifts is almost free of catastrophic
outliers. By retaining this conservative cut, the fraction of
galaxies that are not removed increases according to the depth of
the optical plus IR survey. For the ideal + u + IR case only 5\% of
galaxies are cut and there are no outliers.

We also plot in Fig.\ref{fig:error} a scatter diagram for 20000
galaxies representing the errors estimated by the neural network as
a function of the difference between the photometric and
spectroscopic redshifts. This shows us the reliability of the error
estimate. We can see that the error estimate is relatively good if
there is IR data included whereas it is not optimal with optical
data only; the points with low photometric redshift errors are
concentrated around $z_{spec} - z_{phot} \sim 0$ for optical plus IR
data but not if we have optical data only. In fact when IR data is
included the error estimate overestimates the real error in the
sample. We find that for optical data only, the deeper surveys can
be cleaned by removing the higher error galaxies, however this
becomes harder to do with the shallower surveys. This is mainly
because the magnitude cuts for the sample have been done here at RIZ
of 25.0 and the shallower surveys have large scatters; the large
scatters are dominated by the fainter sources and this biases the
error estimation.

In order to assess whether bands are necessary to obtain good
photometric redshift error estimates with neural networks and be
able to clean the catalogues efficiently we have plotted in
Fig.\ref{fig:hist_err} the unnormalised histograms of the quantity
$(z_{spec}-z_{phot})/z_{err}$. Each galaxy in the mock will have
this ratio; if the error estimate is reliable then the shape of the
histogram for this quantity should be centred around zero and there
should be roughly 68\% of galaxies with $|z_{spec}-z_{phot}|/z_{err}
< 1$. We can see in Fig.\ref{fig:hist_err} that the shallower
optical survey is not able to produce data good enough to produce
reliable photometric redshift errors at a RIZ magnitude cut of 25.0.
In fact it is the range of wavelengths probed that produce a more
reliable photometric redshift error estimate; i.e. by including IR
data the improvement is great, however by including u data or z and
y data it is also possible to obtain good photometric redshift
errors and clean the catalogues as shown previously.

We show the effect of the cleaning in Fig.\ref{fig:error_compare}
where we compare $z_{spec}$ versus $z_{phot}$ scatter plots, for a
range of mocks, before and after the cleaning procedure. We conclude
that IR  or u band data is very important to obtain good photo-z
error estimators with neural networks, when selecting galaxies
fainter than RIZ=25.0 mag.

\begin{figure*}
\begin{center}
\includegraphics[width=18.0cm,angle=0]{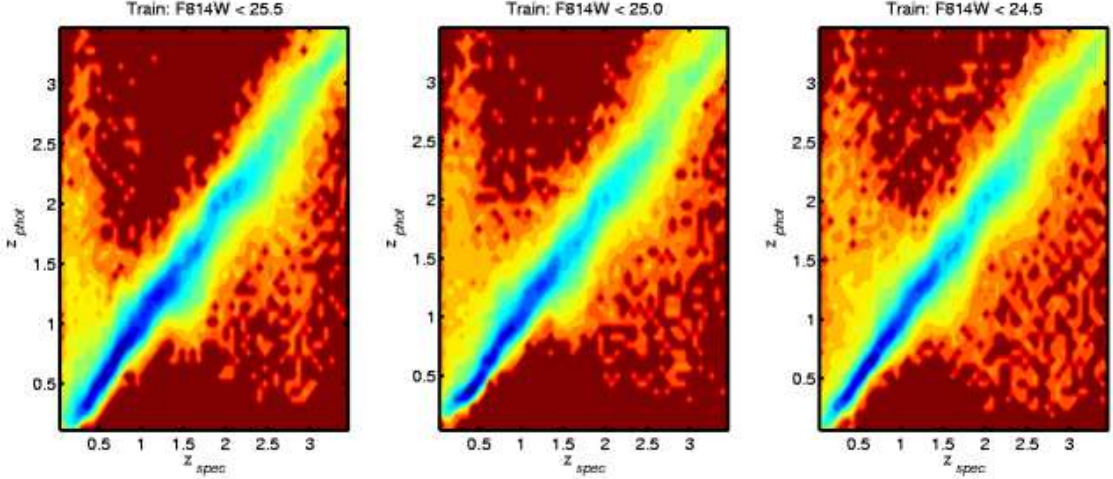}
\caption{Effect of the training set depths in the photometric
redshift quality. We have taken mocks from the survey denominated by
cosmos and we have made cuts in the filter F814W at several
magnitude cuts from 24.5 to 26.5 in steps of 0.5 in magnitude. We
have assumed complete training sets with the same number density of
training galaxies per colour volume for every case shown here. We
can see the bright galaxies with high signal to noise have well
constrained photometric redshifts whereas the noisy galaxies have
almost unconstrained photometric redshifts. The plots are colour
coded and the scale is exponential; each unit corresponds to one
e-fold. \label{fig:cosmos_train}}
\end{center}
\end{figure*}

\subsection{Impact of the source catalogue depth}

In catalogue generation we should always define a certain cut where
we define what an object is. In this paper we have taken a magnitude
cut in a given band to define our catalogues. Here we will show how
this cut influences the quality of the photometric redshifts and the
photometric redshift errors that are associated to each catalogue.
The larger the photometric signal to noise ratio, the better any
code will be able to recover a photometric redshift for that galaxy.
Therefore we can obtain very deep photometry and produce a catalogue
of brighter objects which will have better photometric redshifts and
associated photometric redshift errors or have a catalogue of
fainter sources where the photometric redshifts and their associated
errors are less reliable.

We plot in Fig.\ref{fig:cosmos_dens} the results where we have used
a simulated survey similar to the COSMOS survey as described in
Tab.\ref{tab::depths}. We have used in each subplot only a subset of
the entire set of galaxies we have simulated. The cut was done on
the F814W filter, with depths of 24.5, 25, 25.5, 26 and 26.5. As we
can see the bright galaxies have accurate photometric redshifts,
much more accurate than the faint ones. Particularly, many
catastrophic outliers disappear with the shallow integrations. This
is because some bands are able to distinguish the real redshift of
the galaxy and the catastrophic error, for instance as we have
already mentioned the u band can help remove low redshift galaxies
which are assigned a high photometric redshift. However, a high
signal to noise is needed which is available for the bright galaxies
and not for the faint ones.

We have taken the data used in Fig.\ref{fig:cosmos_dens} and plotted
the scatter as a function of redshift in Fig.\ref{fig:cosmos_sig}.
As we can see there is a definite trend of having a lower scatter if
the sample taken has a high signal to noise ratio for the magnitude
estimates. We can follow from the lower curve to the curve situated
on the top how increasing the depth of a catalogue with the same
data may produce more galaxies with photometric redshifts of worse
quality.

\subsection{Impact of the training set}
\label{ssec:train}

We have assumed so far that the training set used to train the
neural networks is totally representative of the testing set we use
to produce the 2D probability densities and scatter plots to assess
the photometric redshift accuracies. For most cases this requires a
training set which is complete down to a magnitude limit of 25.
Observationally this is a hard task as spectrographs have limited
spectral ranges and the features required for redshift estimation
change in observed wavelength. It should be easier to produce such a
training set down to 24 or 24.5 and in the redshift range 0 to 1.4
\citep{2005A&A...439..845L}.

However the faint end and high redshift range might pose some
problems. One efficient way to get redshift estimates in the range
1.4 to 3 is in the blue optical and UV
\citep{1999ApJ...519....1S,2006astro.ph.12291L} There are many metal
absorption features in this range and Lyman-Alpha comes into the
optical window at $z=1.8$. This is what currently high-z surveys are
doing successfully. Another option is to have a large spectroscopic
redshift survey in the IR which currently does not exist but will be
done with FMOS \citep{2006SPIE.6269E.136D} in the near future. It
should be noted that this is an optimistic scenario as the training
set may not be as wealthy as we have estimated here.


We also have neglected here the contamination of other unusual
objects that might be introduced in the sample. For instance,
low-luminosity Seyfert-1 galaxies will have strong lines and a
different continuum shape than the usual galaxies considered. These
could, in reality, account for roughly a XX per cent of extra
objects in the training and testing sets. We note however that a
neural network is a Bayesian object therefore, if the colours of
these unusual objects are totally different from the colours of the
other galaxies then the network is flexible enough not to interfere
on the training of the usual galaxies and the same result will be
found. If the unusual objects have similar colours to the usual
galaxies then the fact that the network is Bayesian will be to our
advantage. Given that the contamination is only small, the weight of
these unusual objects will be diluted as they are less
representative. We therefore argue that a small contamination of
unusual objects whichever their colour will not affect the photo-z
of usual galaxies considerably.

We stress here that the errors which will be associated to training
sets arise from two terms, one is the square root of the number of
spectra $N_s$ available in our analysis. The second is the rms sigma
as a function of redshift for that group of galaxies. Weak
gravitational lensing is sensitive to the error on the mean of the
redshift for galaxies, which is dependent on the quantity
$\sigma^2(z)/N_s$. This error is complex to analyse as $\sigma(z)$
depends on the photometric bands as well as the method used for
photometric redshift estimation. An analysis of this is made in
Sec.\ref{wl_fom}.

\begin{figure}
\begin{center}
\includegraphics[width=8.5cm,angle=0]{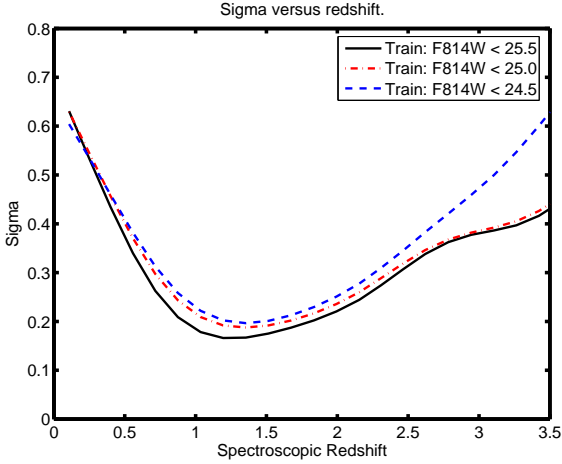}
\caption{The scatter of $z_{phot} - z_{spec}$
as a function of redshift for samples
with different training sets. Same data as in Fig.\ref{fig:cosmos_train}.
As we can see
the sample which has been trained with an incomplete training set
has worse photometric redshifts with many more outliers and a larger scatter,
however the decrement in accuracy is not extremely large.
\label{fig:train_sig}}
\end{center}
\end{figure}

To assess the impact of an incomplete training on the accuracy of
the photometric redshifts produced we have used training sets with a
brighter magnitude cut and estimated photometric redshifts in a
fainter sample. We have maintained the density of training galaxies
per unit of colour volume the same in all the runs. We have chosen a
cut of 25.5 in the F814W band and trained the neural networks with
sets cut at 25.5, 25 and 24.5 in the same band. The probability
density plots are shown in Fig.\ref{fig:cosmos_train} and the
scatter as a function of redshift in Fig.\ref{fig:train_sig}. As we
can see an incomplete training produces slightly worse photometric
redshifts. If we assume that we can extrapolate from the colours of
the brighter objects, having only a training set complete to 0.5
magnitudes brighter degrades the photometric redshifts by about
20\%. There are however, as shown in Fig.\ref{fig:cosmos_train}, a
much larger number of catastrophic outliers which are mainly the
galaxies with no representatives in the training set.

\section{Colour \& Type analysis}
\label{cor_type}

In this section we attempt to analyse and establish which types of galaxies
are producing large catastrophic outliers in the
photometric redshift analysis. This is important to assess which galaxy
properties are introducing the larger errors in our analysis, this would
allow us to have a greater understanding of how to reduce systematic
effects due to photometric redshifts and have a comprehensive understanding
of how different bands can help the photometric redshift analysis.

For this purpose we choose two different catalogues to perform our
colour and type analysis. We choose first a catalogue obtained from
DES like exposure times that has been cut in magnitude at $r <
24.0$, which is the depth of the photometry in the optical. We also
choose another optical catalogue with an RIZ depth of 25.0 and with
the specifications of the ideal optical survey we have chosen in
Sec.\ref{sec:impact_bands}.

\begin{figure*}
\begin{center}
\begin{minipage}[c]{1.00\textwidth}
\centering
\includegraphics[width=7.2cm,angle=0]{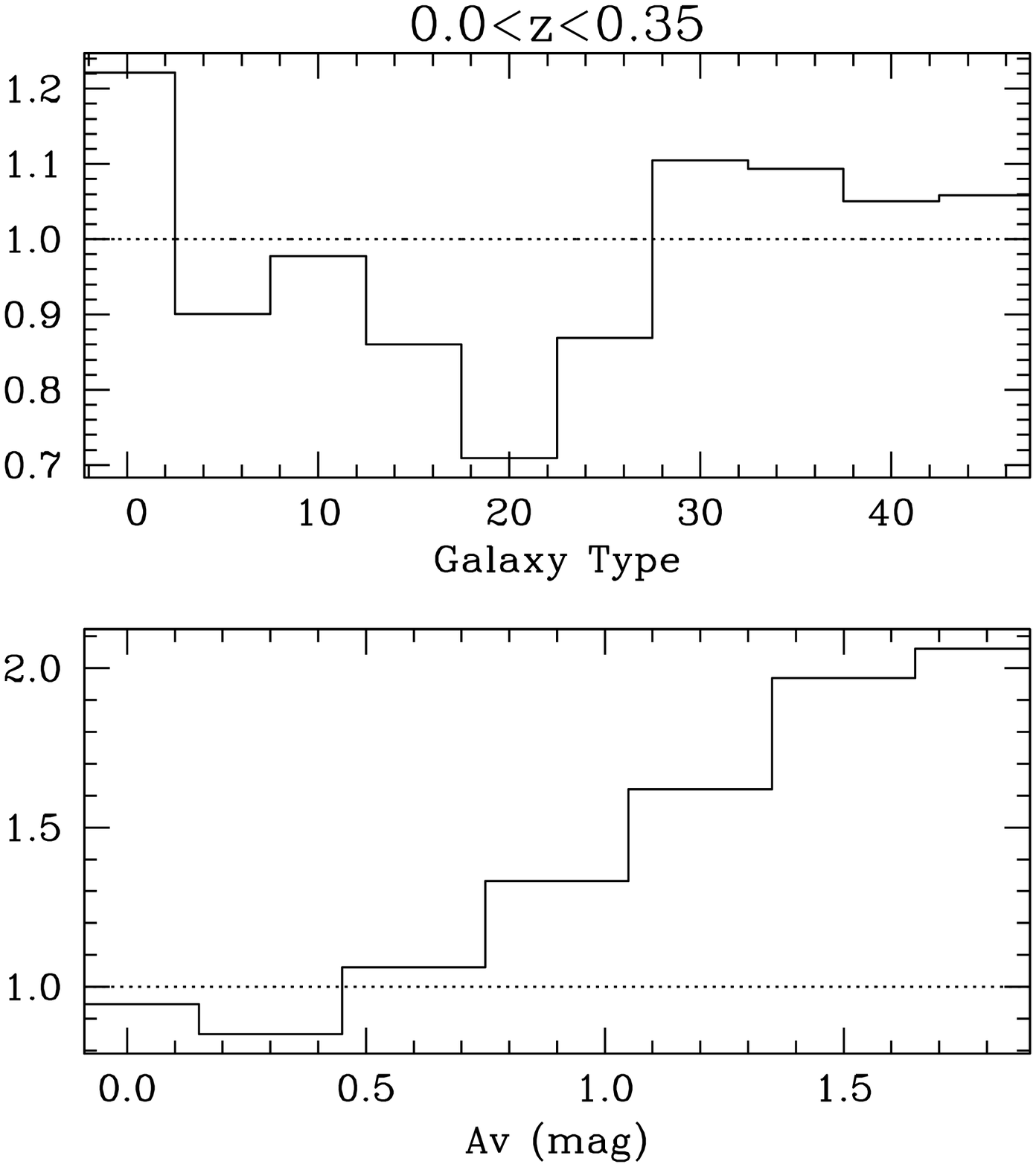}
\includegraphics[width=7.2cm,angle=0]{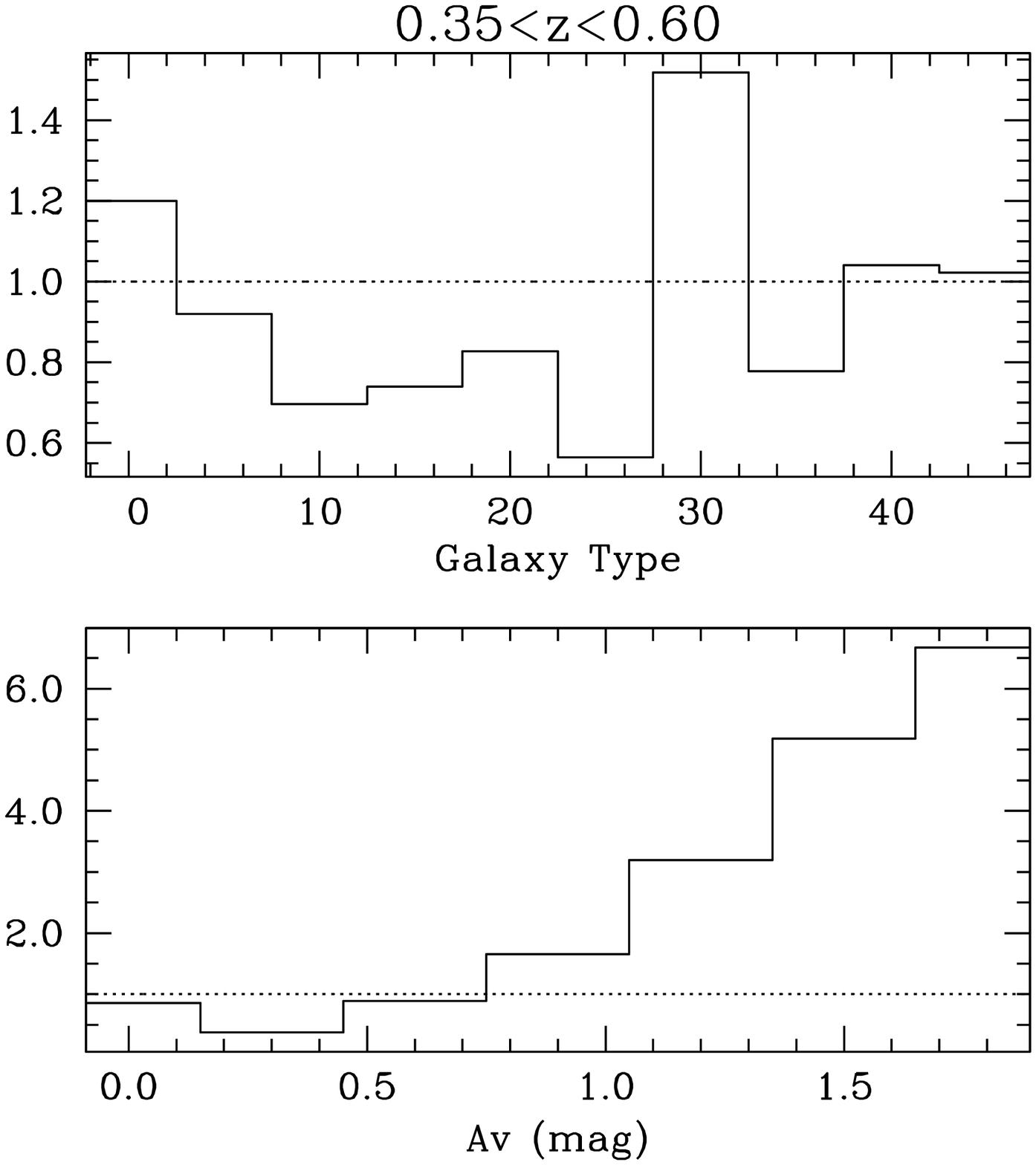}
\includegraphics[width=7.2cm,angle=0]{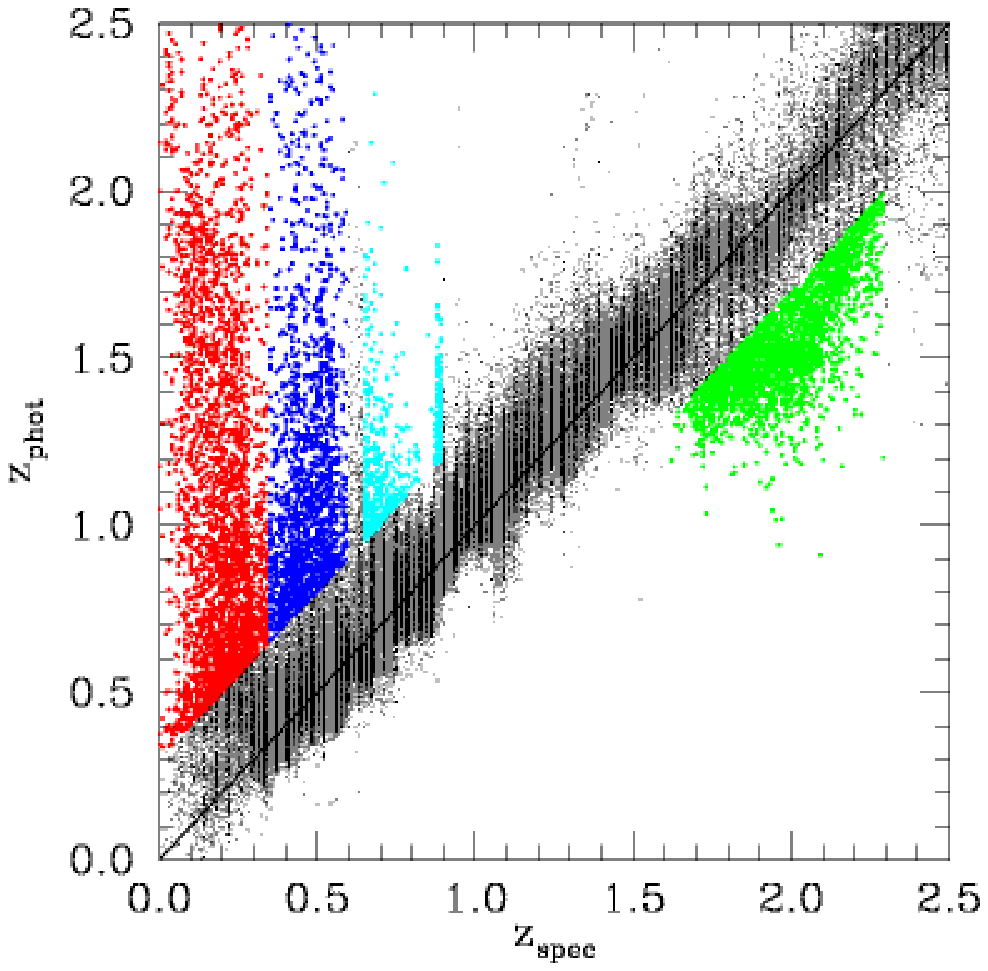}
\end{minipage}
\begin{minipage}[c]{1.00\textwidth}
\centering
\includegraphics[width=7.2cm,angle=0]{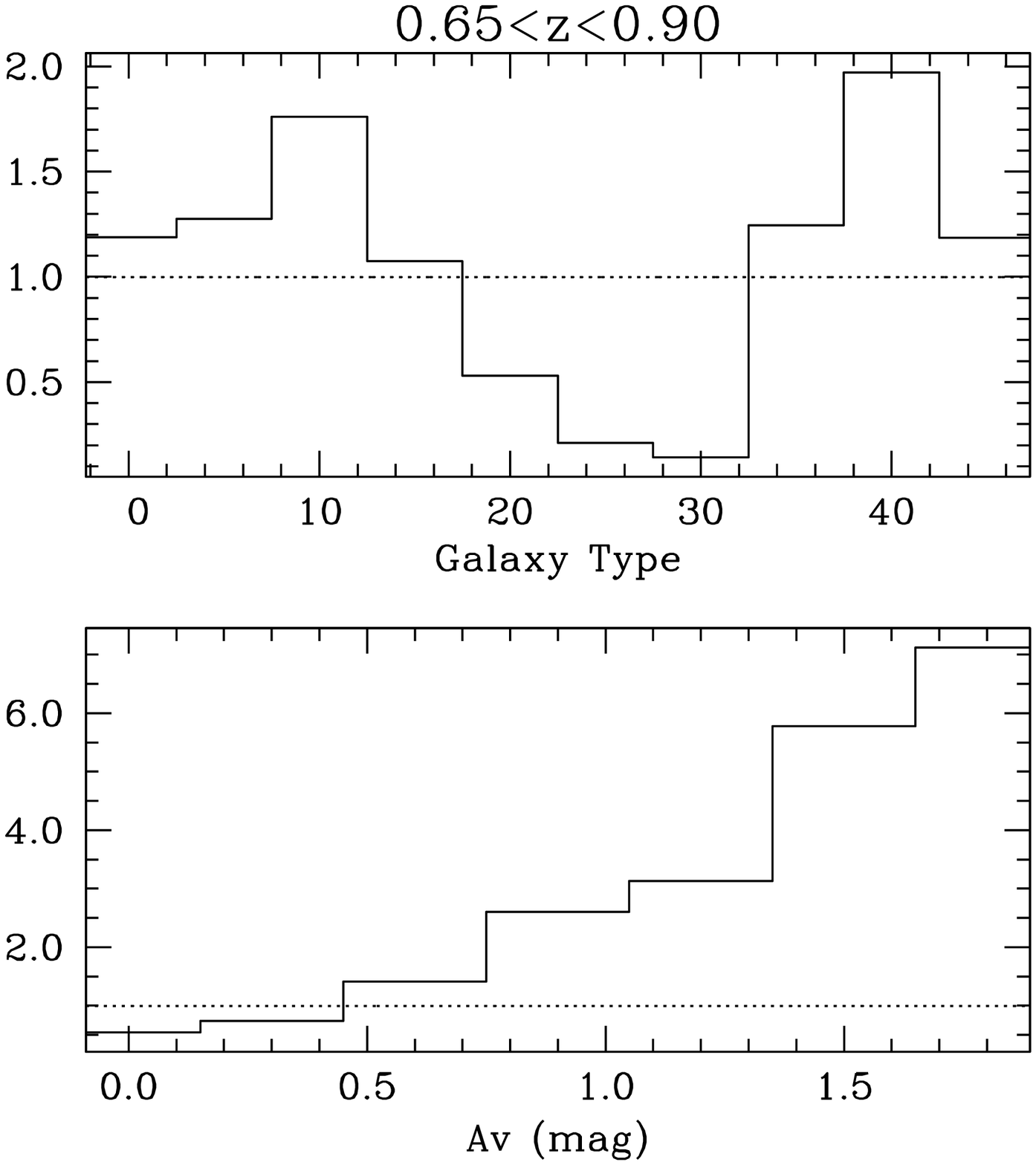}
\includegraphics[width=7.2cm,angle=0]{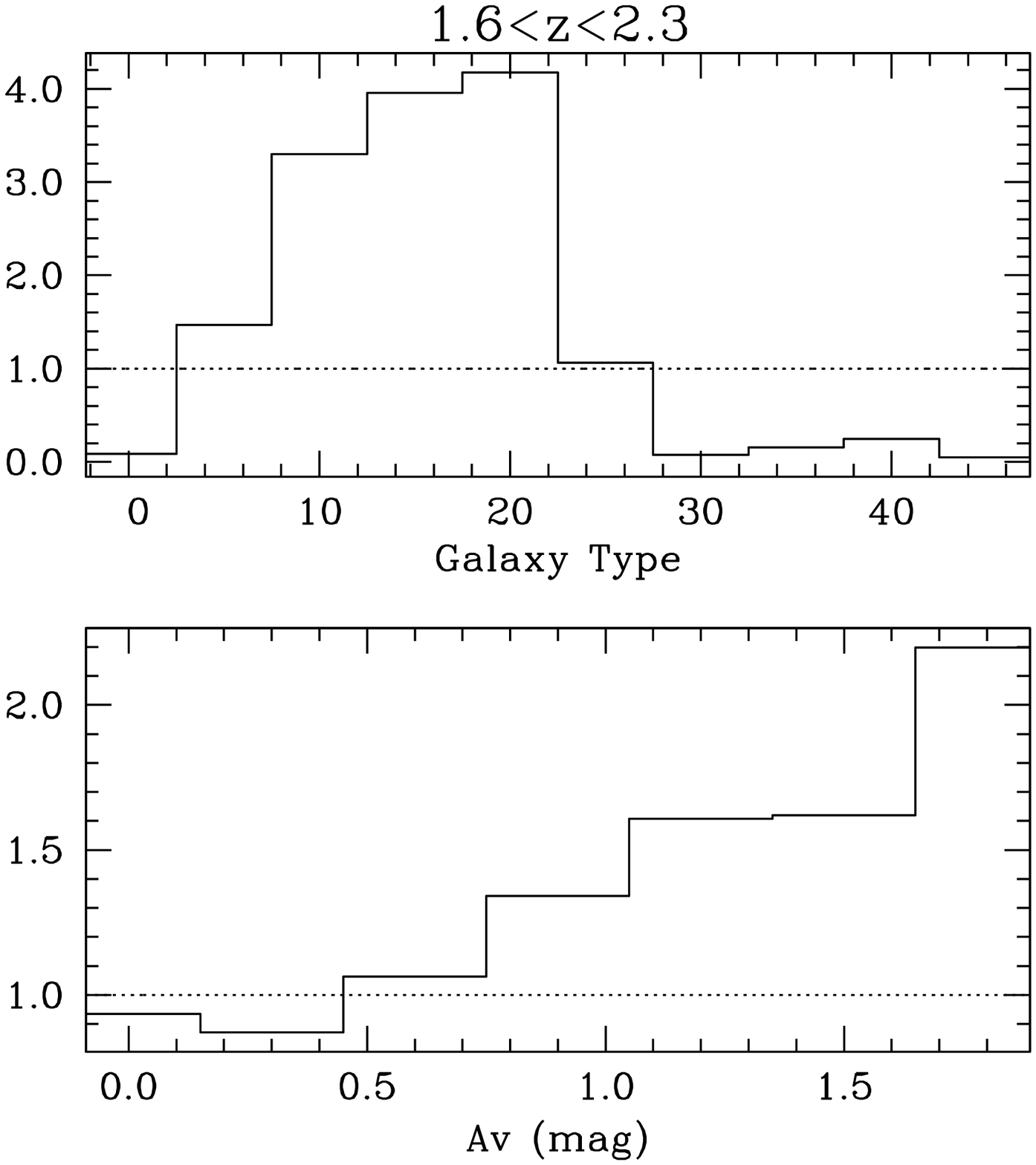}
\end{minipage}
\caption{Mock data with the
survey we have labelled by ideal optical survey without u band
photometry. We find that despite the difference in the magnitude
limit of the survey the regions with a high concentration of
outliers contain similar types of galaxies which are mainly reddened
galaxies and/or starburst galaxies. The main difference is that in
the region 1.6 to 2.3 in redshift the outliers have a low fraction
of elliptical galaxies. This is due to the high exposure times in
the y and z bands chosen in this mock ideal survey.
\label{fig:zone2_ideal}}
\end{center}
\end{figure*}

We show in the centre of Fig.\ref{fig:zone2_ideal} the scatter plot
for the photometric redshifts as a function of the spectroscopic
redshifts for the fourth case considered. From this figure we have selected
four spectroscopic redshift bins (0.0--0.35, 0.35--0.6, 0.65--0.9
and 1.6--2.3) which contain large numbers of ouliers. From these
regions we selected all points with $|z_{phot} - z_{spec}|
> 0.3$ on one of the sides of the  $z_{phot} = z_{spec}$ curve.


We plot on the four panels above and below the scatter diagram
the relative histograms of the populations within the regions selected
relative to the average population of galaxies considered in the magnitude cut.
This means that if the histograms are above one this galaxy type or
$A_v$ (the extinction in magnitudes in V band)
is dominant in the region selected, whereas if the histogram is below one
then the population in the region is sub-dominant.


\begin{figure}
\begin{center}
\includegraphics[width=8.5cm,angle=0]{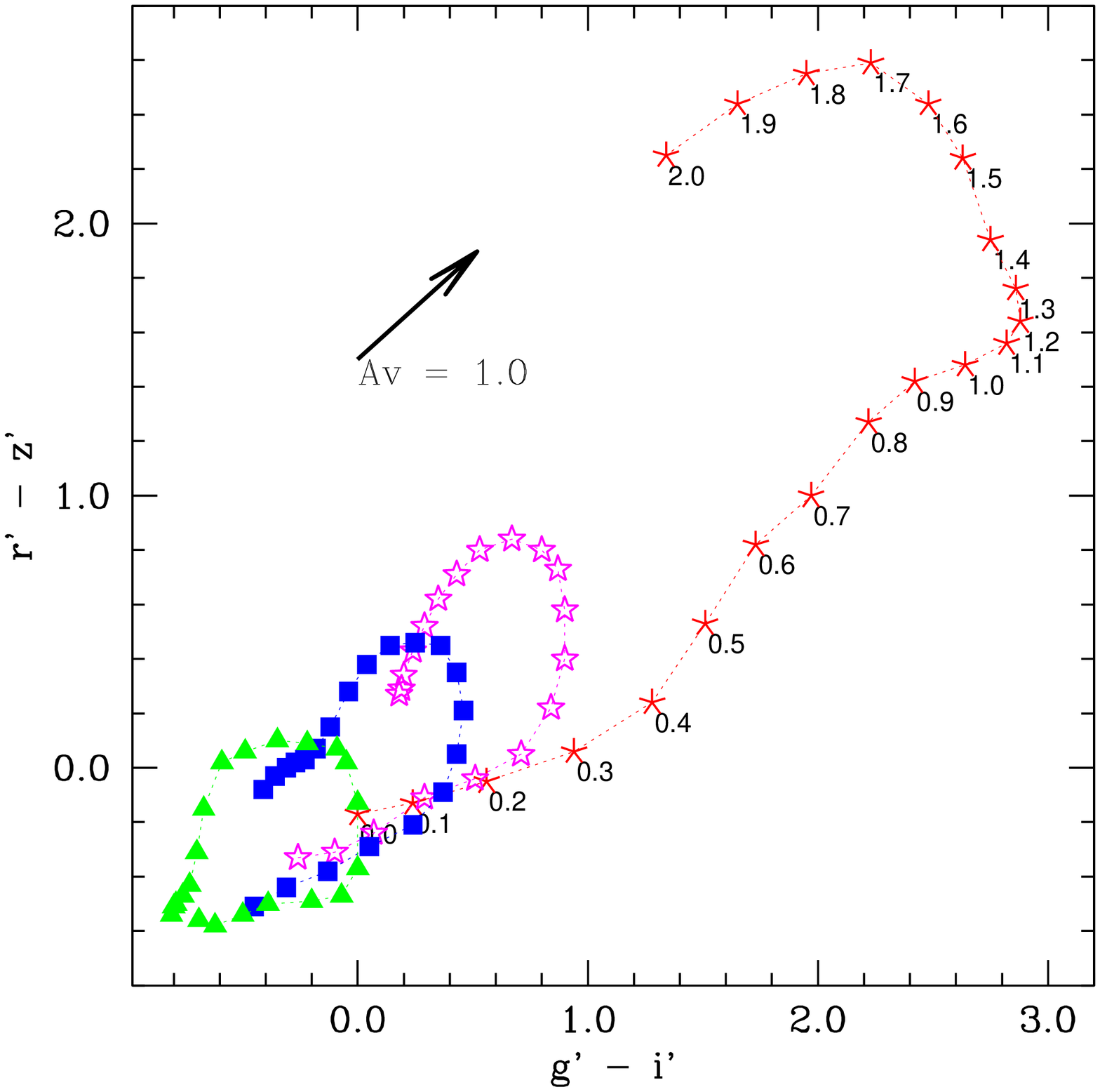}
\caption{Colour-Colour (g-i $\times$ r-z)  diagram of the different
galaxy types with redshifts ranging from 0 to 2.
The red points represent
elliptical galaxies. The bluest point correspond to the colours of
an elliptical galaxy at $z=0$, the other points are the same galaxy
further along in redshift up to $z=2.0$. Pink, dark blue and  green,
light blue points represent Sbc, Sdc and Irr types of galaxies.
We have also plotted the vector in this plane which corresponds to the shift
in colour for a reddening of $A_v = 1.0$. As we can see the shift in the
colour-colour plane corresponding to redshifting the galaxy is degenerate in
some cases having a reddened galaxy. In other words a faint reddened
galaxy at low redshift has very similar colours to a brighter galaxy at
higher redshift with no reddening. Similarly a very high redshift galaxy
which is reddened has the same colours to an intermediate
redshift galaxy which is not reddened.
\label{fig:cor_cor}}
\end{center}
\end{figure}

We can see that in the first region, most galaxies that are
scattered towards higher photometric redshifts have a high $A_v$ in
our photometric catalogue. This means that galaxies which are not
reddened have relatively good photometric redshifts whereas galaxies
which are heavily reddened are scattered up towards high redshift.
We can see from the second region chosen that the same occurs there,
however this happens for different types of galaxies. The same
occurs for the fourth region chosen, above a redshift of 1.6, but
here the galaxies which are severely reddened are scattered towards
lower redshifts. The third region we have selected is heavily
populated by galaxies with types close to 50 which corresponds in
our notation to very blue and young starburst galaxies which are
naturally hard to get photometric redshifts for because of their
relatively featureless continuum. We conclude from these graphs that
the main source for catastrophic photometric redshifts for an
optical survey arises from heavily reddened galaxies
as well as very blue starburst galaxies.

We have performed the same analysis with similar regions for 
$r<24$. The results are very
similar to the results obtained for a cut one magnitude fainter. The
main differences are that in region four with redshifts larger than
1.6 the late type galaxies with type close to zero are well
constrained by the optical data given the very deep exposure times
in the z and y bands. Furthermore the scatter in that region is
significantly lower. However, the less well constrained objects have
still same properties of being either blue starburst or heavily
reddened objects.

There is a big degeneracy
between redshift and reddening for some redshift ranges. We plot for
different galaxy types the colour in the r-z and g-i plane for
galaxies at different redshifts. We also plot the vector which
corresponds to the average shift in colour for a reddening of $A_v =
1.0$. Clearly galaxies at low redshift which are reddened have very
similar colours to galaxies at higher redshifts which are not
reddened. We can see with this explanation why galaxies are
scattered upwards in the $z_{spec} - z_{phot}$ plane. We have used a
neural network to obtain photometric redshifts; given that the
training set we have chosen is representative there will be more
galaxies at redshift $z \sim 1$ than at low redshift. We argue that
the inclusion of IR data helps the distinction between reddened
galaxies and galaxies with a low $A_v$ given the different
extinction as a function of wavelength which is low at IR bands. The
case which we have chosen with an ideal + u survey can also
distinguish between galaxies with a high and low $A_v$; this is
because at very high exposures for the u band, a detection in u
would allow us to know whether the galaxy is reddened. This is
because of the high extinction in the u band.



\section{Weak Lensing tomography: the dependence of the Dark Energy parameter estimation
on the Photo-z accuracy}
\label{wl_fom}

\subsection{A qualitative treatment of the WL-photoz cross talk}

One important application of photometric redshifts is for the analysis
of weak lensing tomography \citep{1999ApJ...522L..21H} ,
as e.g. in the DUNE experiment.  The idea
is to slice a lensing survey into photometric redshift bins and to analyse the
cosmic shear for high signal-to-noise galaxy images  in each
photo-z bin, e.g. as discussed by \citet{2006ApJ...636...21M}, hereafter
MHH and by \citet{2006astro.ph.10127A}, hereafter AR.

Specifically, we can divide the galaxy sample into photo-z bins and
examine the effect on the derived cosmological parameters due to
uncertainties in the photo-zs.  Consider a distribution of sources
selected from a photometric redshift bin which results in a more
complicated (not necessarily Gaussian) distribution with
respect to true (spectroscopic) redshift, with mean redshift ${\bar
z}$ and variance:
\begin{equation}
 \mu_2=\left<(z_{spec}-\bar z)^2\right>.
\label{eq:mu2}
\end{equation}

Assuming Poisson statistics we can predict the variance in the mean
redshift ${\bar z}$  given $N_{spec}$
spectroscopic redshifts  associated with that photo-z bin:

\begin{equation}
(\delta z)^2  \equiv
rms^2({\bar z}) = \mu_2/N_s\;.
\label{eq:var1}
\end{equation}

We can now model crudely the the uncertainty in deriving the
constant Dark
Energy parameter $w=P/\rho$ from WL if the only uncertainty is due to
photo-z errors:
\begin{equation}
|( \delta w) / w| = a  (\delta z)/{\bar z} \;.
\label{eq:dw}
\end{equation}

The `fudge factor' $a$ can be estimated from detailed modelling of the WL power spectrum.
For example, if we set all other parameters to be known in the scaling relation given
in \citet[][Eq.24]{2006MNRAS.366..101H}
we find that $ {\bar z}^{1.6}  \propto |w|^{0.31}$
and hence $a=5.2$. This value $a \approx 5$ can also be
justified qualitatively by examining the sensitivity of cosmological
distance and the linear growth to variations in $w$
\citep{2006Msngr.125...48P}.

We can now combine the last two equations to give:

\begin{equation}
|( \delta w) / w| = (a /{\bar z})  {\sqrt{\mu_2/N_s}}\;.
\label{eq:dw-nspec}
\end{equation}

For example, for a desired fractional error of 1\% on $w$, $a=5$ ,
${\bar z} = 1$ and $\mu_2 = 0.06$ (derived from our mocks and ANNz
averaged over a range of proposed optical and IR surveys) we find that
$N_s \approx 15,000$ spectroscopic redshifts are required for that bin,
or for say 10 bins a total of $150,000$ spectroscopic redshifts.

Our back-of-the-envelope calculation helps us to understand the link
between the Dark Energy parameters, the photometric redshift
performance and the number of spectroscopic redshifts.  However,
being derived only for a constant $w$ and with other cosmological
parameters fixed, it somewhat under-predicts our detailed
calculation below which is done for a 2-parameter equation of state
after marginalisation over other parameters.

More generally \citet{2006astro.ph.10127A} have considered the way the
Figure of Merit (FOM)
for the Dark Energy parameters $(w_0, w_a)$ is affected by photo-z
errors alongside the effect of the sky coverage and depth, the shear
measurements systematic and uncertainties in the non-linear power
spectrum predictions.  Their scaling relations roughly agree with our
detailed FOM calculations described below.

\subsection{Dark Energy Figure of Merit (FOM) calculation}

We summarise below the ingredients for the FOM calculation for Dark
Energy parameters from weak lensing.
The predicted angular power spectrum $C_{ij}(l)$ between redshift bins
$i$ and $j$ depends on the 3-D matter power spectrum $P(k=l/r)$ and on the
radial window functions $W_i(r)$ and $W_j(r)$
of bins $(i,j)$:
\begin{equation}
C_{ij}(l) = \int_0^{r_{H}} dr r^{-2} W_i(r) W_j(r) P(l/r; r)+
\delta_{ij} \sigma_e^2/{\bar n}_i\;,
\label{eq:clij2}
\end{equation}
where r is the comoving distance and $r_H$ is the Universe horizon.
The last term is the `shot noise'  due to $\sigma_e$, the intrinsic
ellipticity noise for the galaxy sample,
and  ${\bar n}_i$, the total number of galaxies in
the radial slice. The case $i=j$ gives the auto-correlation of bin
$i$.  The window function $W_i(r)$ depends on cosmological
parameters and on the redshift distribution $p_{i}(z)$ of the source
galaxies redshift slice, normalised such that ${\bar n}_i = \int dz
p_i(z)$ is the total number of galaxies in the slice and the integral is
over the slice's radial boundaries. For a comprehensive list of
detailed equations in weak-lensing correlations
we refer the reader to \citep{2007arXiv0705.0166B}.

Ideally, we would like to know the $p_i(z)$ exactly from a
spectroscopic survey, where the redshift is derived from the spectrum
of each galaxy. In reality, deep wide field surveys, such as  DUNE, will
only provide us with multi-band imaging data which will allow us to
derive photometric redshifts based on templates and/or spectroscopic
training sets.  We can relate the probabilities for the true redshift
$z_{spec}$ and the photometric redshift $z_{phot}$ by the Bayesian rule of
conditional probability:

\begin{equation}
p(z_{spec}, z_{phot}) = p(z_{spec} |  z_{phot}) p(z_{phot}) =
p(z_{phot} | z_{spec}) p(z_{spec})
\label{eq:bayes}
\end{equation}

Consider now a sharp cut in a photo-z bin $i$ , i.e.  we select only
those galaxies in the range $z_{phot}(i) < z _{phot} < z_{phot}(i+1)$.  We can
write the probability for the true redshift distribution resulting from
the photo-z slice as:
\begin{equation}
p_i(z_{spec}) = \int_{z_{phot}(i)}^{z_{phot}(i+1)} p(z_{phot},z_{spec}) \; dz_{phot}
\label{eq:pizt1}
\end{equation}
Typically $p_i(z_{spec})$ would have a wide spread, {\it not} in the
form of a Gaussian, as a-symmetric tails are present due the photo-z
catastrophic errors.

We have now two options.  One is to parameterise $p_i(z_{spec})$
directly based on the projection of the photo-z scatter diagram
$p_i(z_{spec}, z_{phot})$, derived from a spectroscopic training set
or mock catalogues.  The second option is to model it using eq.
(\ref{eq:bayes}):
\begin{equation}
p_i(z_{spec}) = \int_{z_{phot}(i)}^{z_{phot}(i+1)} p(z_{spec}) p(z_{phot}|z_{spec}) \; dz_{phot},
\label{eq:pizt2}
\end{equation}
where $p(z_{spec})$ is the overall galaxy redshift distribution and $p(z_{phot}|z_{spec})$ can be modelled, somewhat ad-hoc, as a Gaussian (e.g. MHH, AR).

Here we prefer the first option, i.e.
we  use the actual distributions of spectroscopic redshifts coming out of the simulations
for the analysis.

The uncertainties in the shapes of the probability distribution
function also have to be taken into account, due to the finite number,
$N_s$, of the spectroscopic redshifts per bin in the training set.
We have assumed an uncertainty on the mean and on the variance of the
distributions we have from the photo-z simulations for each redshift shell
in our analysis. As explained in the previous sub-section,
from Poisson statistic we can predict the variance in the mean
redshift of the bin:

\begin{equation}
rms^2({\bar z}) = \mu_2(z_{spec})/N_s
\label{eq:var2}
\end{equation}

\noindent where
 $\mu_k(z_{spec})=\left<(z_{spec}-\bar z)^k\right>$ for a given
photometric redshift bin. Similarly  for the variance in the variance

\begin{equation}
rms^2({\mu_2}) = \frac{N_s-1}{N_s^3}[(N_s-1)\mu_4-(N_s-3)\mu_2^2]
\simeq   \frac{\mu_4-\mu_2^2}{N_s}.
\label{eq:var3}
\end{equation}

\begin{table}
  \begin{center}

    \begin{tabular}{|l|c|c|}
      \hline          & + RIZ     & + RIZ + IR  \\
      \hline
      DES-like        & 72/83.6   & 120/124.0  \\
     \hline
      Pan-4-like      & 80/87.3   & 132/130.8 \\
     \hline
      LSST-like       & 116/112.0 & 156/148.0  \\
     \hline
      Ideal-like      & 136/130.5 & 168/160.4 \\
     \hline
      Ideal + u-like  & 164/151.0 & 168/162.3  \\
     \hline

    \end{tabular}    \vspace{2mm}
  \end{center}
  \caption{The FOM for the survey configurations we consider with and
without IR data. The FOM results are for a lensing survey only and do
not include any other prior such as CMB. We note here that this is not a
straight comparison for the different surveys as we only consider galaxies
which will be observed by DUNE. For instance the real DES survey will be only
over a smaller area of the sky and a LSST survey would be much deeper and
hence have many more galaxies usable for lensing. The numbers on the
left are the numbers computed in this work using the full distribution of
redshifts. The numbers on the right are FOM values using the fitting formula
given by AR.}
  \label{tbl:foms}
\end{table}

One can marginalise over both these uncertainties. The dependence on
the number of spectroscopic redshifts is really an indirect
expression of the scatter in ${\bar z}$ and in $\mu_2$.

The Dark Energy equation of state is commonly written as: $w(a) =
w_0  + (1-a)w_a$. One can also define $a_p$, the pivot value at
which the uncertainty in $w(a)$ is the minimum. Accordingly we
define $w_p = w_0 + (1-a_p) w_a$. Armed with this parameterisation
we can now apply the Fisher matrix formalism. We define
FOM\footnote{The Dark Energy Task Force report
\citep{2006astro.ph..9591A} defined The Figure of Merit slightly
differently as the reciprocal area in the $w_0-w_a$ plane that
encloses 95\% CL region. It differs to our FOM by a constant
factor.} as:
\begin{equation}
{\rm FOM}=\frac{1}{({\rm det} F_{ij}^{-1})^{1/2}}
=\frac{1}{\delta w_p \delta w_a}\;
\label{FOM}
\end{equation}
where $i$ and $j$ denote the elements of the covariance matrix ($F^{-1}$)
that contain the equation of state parameters $(w_p,w_a)$) and $\delta
w_p$ and $\delta w_p$ are the 68\% errors on $w_p$ and $w_a$.

In our Fisher matrix analysis we vary the following
cosmological parameters: the dark matter content $\Omega_m$, the dark energy
parameters $w_0$, $w_a$, the Hubble constant divided by 100 $h$, the
amplitude of fluctuations at $z=0$ $\sigma_8$, the baryonic matter content
$\Omega_b$ and the scalar index of the primordial spectrum $n$
where the fiducial values are 0.3, -0.95, 0.0, 0.7,
0.8, 0.045 and 1.0 respectively. We assume spacial flatness.
As we have already stated we have
also assumed that there is an uncertainty over the mean and variance for
each of the redshift bins independently. Hence we have two extra nuisance
parameters that we will marginalise over for each photometric redshift bin.

When performing lensing measurements, cuts are made on the detected
galaxies.  The two most significant cuts are in flux, where galaxies
below a certain magnitude threshold are not used and a size cut
where small galaxies are rejected because their shape is hard to
measure. In this work we have not included a size cut since this
would require detailed image simulations which is beyond the scope
of this paper. However the effect of this cut is important since
this will remove high redshift galaxies from our sample.  To mimic
the effect of these cuts we randomly sample the galaxies in our mock
and apply a selection so that our galaxy population has a more
realistic PDF.  Specifically we impose a distribution ($P(z) \propto
z^2 \exp (-(z/z_0)^{1.5})$), where $z_0$ is set by the median
redshift of the galaxies ($z_0=z_m/1.412$), which we assume to be
$z_m = 0.9$.


We have performed all the analysis with five photometric redshift bins and also checked that having
more photometric redshift bins did not improve considerably on the
FOM. We have chosen the photometric redshift bins so that
the number of galaxies in each bin was the same. The total number of
galaxies considered was 35 ${\rm gal}/{\rm arcmin}^2$ and we have considered
a survey covering 20000 ${\rm deg}^2$.

\begin{figure}
\begin{center}
\includegraphics[width=8.0cm,angle=0]{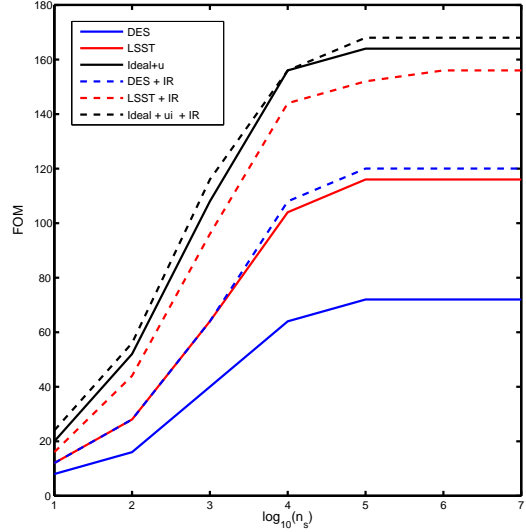}
\caption{The FOM as a function of the number of spectroscopic
redshifts (per bin) used to calibrate the photometric results. We can see that
the FOM began to level off at $10^4$ and having
more than $10^5$ spectroscopic redshifts does not increase the
FOM thus meaning that the photo-z are well enough calibrated. We conclude that
for a DUNE-like survey we will need around $10^5$ objects in each redshift bin
(here we assumed 5 bins) to calibrate the photometric redshift. Around $10^4$
objects might be sufficient, however the use of Eqn.\ref{eq:var2}
does not take account
of outliers or the non-Gaussianity of the distribution
and a number closer to $10^5$ will be necessary.
Furthermore
these redshifts must be representative of the sample and therefore most of
them should be of galaxies in the faint end of the catalogue, i.e. fainter
than RIZ of 24.
\label{fom::ns}}
\end{center}
\end{figure}

\subsection{FOM results}


We now investigate the role of the IR filters in the FOM prediction.
We can clearly see that the role of the IR filter depends on the
quality of the optical data available. We found before that the
scatter plots for the photo-z benefited significantly by the
inclusion of IR data. Here we find that if we have poor optical data
with a depth similar or shallower than the target depth of 25 in the
RIZ filter, then the IR bands can improve the FOM by a factor of
1.7. However if the optical data available is deeper than the target
depth of 25 in the RIZ filter, especially in the z and y bands then
this FOM improvement\footnote{In comparison the DEFT report
\citep{2006astro.ph..9591A} suggests that a new generation of
surveys should improve the FOM by a factor of 3. Here the
improvements with IR are a factor of 1.3 to 1.7 within the same
survey generation.} is reduced to 1.3 and if very deep u band data
are available then the improvement is minimal for the purposes of
Dark Energy determination.

All the numbers above assume an infinite number of spectroscopic redshifts
 are available to calibrate the photometric data. In practise only
a finite number is necessary in order to calibrate the data well
enough so that there is no degradation of the FOM. In order to asses
this we have performed Fisher matrix calculations where we have
prior knowledge given by $N_s$, the number of spectroscopic
galaxies. We show the results in Fig.\ref{fom::ns}. As we can see if
we have around $10^5$ galaxies in each of the photometric redshift
bins that we have assumed we have very little degradation of the
FOM. We therefore conclude that we will require that many
spectroscopic redshifts to calibrate the photometric sample well
enough. Obtaining this many spectra is not a daunting process.
However, we emphasise that most of these spectroscopic redshifts
should be of galaxies in the faint end of the sample which will be
the most numerous. A sample of $10^5$ galaxies at high brightness
would not be suitable. As we discussed in Sec.\ref{ssec:train} this
spectroscopic sample is currently not available, however, with IR
surveys probing galaxies at a redshift of 2 combined with surveys
such as zCOSMOS the prospects of having such a sample in the next
decade is not unfeasible. However, there is need for more detailed
study to assess whether this sample will be adequate in terms of
completeness to be able to calibrate photometric redshifts.

We have listed in Tab.\ref{tbl:foms} the results for each survey and
compared them with results given by the fitting formula from AR. We
have used the values of sigma 68 as the values for the scatter for
this distribution and we defined the fraction of outliers as the
galaxies which were situated 3$\sigma_{68}$ or more away from the
mean photometric redshift of that bin. We can see from the results
of the table that the full calculations assuming the full PDF give
very similar results to the fitting formula from AR.

\subsection{Other considerations: systematic effects.}

\begin{figure}
\begin{center}
\includegraphics[width=8.0cm,angle=0]{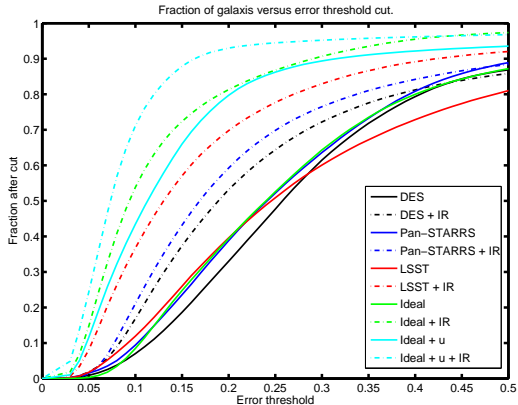}
\caption{The fraction of galaxies removed from each of the
surveys considered if we apply a cut in the photometric redshift error.
As we can see the inclusion of the IR bands increases significantly the
accuracy of the error determination in our analysis. We also note that
among the mock sets with simulated optical data only, the one with u band
data is also reliable in obtaining a reliable error estimate. Furthermore
as it is argued it is possible to clean photo-z catalogues removing
outliers without decreasing significantly the FOM for Dark Energy parameters.
\label{fig:cut}}
\end{center}
\end{figure}

We have assessed the impact on the statistical errors that different
photometric redshift distributions would have on constraining the
Dark Energy parameters. In a weak lensing survey, however, there
will be further systematic barriers which would not allow this
statistical limit to be reached.

One of the important effects that will have to be removed or modelled
and fitted for is the effect of intrinsic alignments for close-by
galaxies. The intrinsic-intrinsic (II) power spectrum introduced by
galaxy intrinsic alignments can be written as

\begin{equation}
C^{I}_{ij}(l) = \int_0^{r_{H}} dr r^{-2} p_i(r) p_j(r) P_{II}(l/r; r),
\label{eq:clij1}
\end{equation}

\begin{figure}
\begin{center}
\includegraphics[width=8.0cm,angle=0]{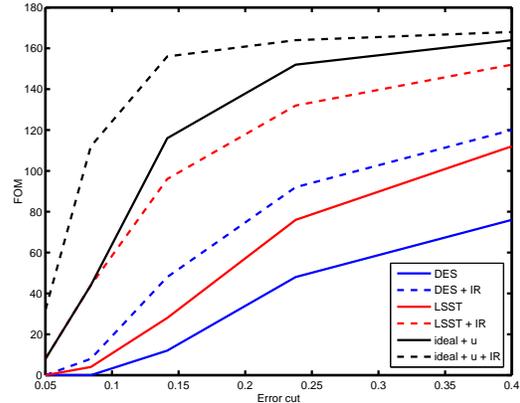}
\caption{The FOM as a function of the photometric redshift
error cut. It can be noted that if we have
a survey with deep IR bands or a survey with deep u band photometry then
performing a cut to remove outliers does not decrease the FOM significantly.
This is clearly a function of how deep the optical survey is. We can see
in Fig.\ref{fig:clean}
the quantity of outliers left on the mocks with a cut of 0.3.
\label{fig:FOMcut}}
\end{center}
\end{figure}

\noindent where $p_i$ is the redshift distribution for the
galaxies in bin i and $P_{II}$ is the intrinsic alignment power spectrum.
We can see that if  $p_i$ and $p_j$ are independent the
contamination of the II term is minimal. Therefore weak lensing
surveys would require a low overlap of galaxies between different bins.
Several ideas have been proposed recently to improve the
performance of the WL analysis.  \citet{2006astro.ph..9338J}
suggested a `colour tomography' to bin the galaxy data in colour space
where galaxies have small overlap in the $p_i$, rather than to generate
a photo-z catalogue and then bin it.

It is beyond the scope of this paper to do a complete analysis of
how these systematic effects will hinder the FOM for the Dark Energy
parameters. We refer the reader to \citep{2007arXiv0705.0166B} for a
more detailed analysis of this. We have assessed to what extent a
detailed analysis of photometric redshift errors would help decrease
the contamination of the II term in weak lensing. We have run Fisher
matrices for catalogues which were cut in increasing redshift error.
That is to say that galaxies that are found to have high error
estimates are removed from the sample, this can be seen in
Fig.\ref{fig:cut}.
We plot the fraction of galaxies left in this sample.
Consequentially, there are less systematic errors as there are fewer
outliers in the sample. We can see from Fig.\ref{fig:clean} how this
affects the number of outliers in the photo-z analysis for a given
cut in the estimated error.

We plot in Fig.\ref{fig:FOMcut} how the FOM is decreased by a
photo-z error cut for some of the surveys considered. We can see
clearly that some surveys can have a large cut in the photometric
redshift error and their FOM remains almost unchanged. This clearly
means that we will be removing systematic errors due to the overlap
of the photometric redshift bins but not hitting the FOM by removing
galaxies which introduce relevant information in the determination
of Dark Energy. Furthermore, if there is a need to model a
galaxy-intrinsic (GI) alignment contribution to the cosmic shear
signal, this will be more reliable if we obtain a sample with only
the reliable photometric redshifts and which does not decrease the
FOM significantly compared to the full sample.

\section{Conclusions}

In this work we have looked at the role of optical and near-IR
photometry in the context of weak lensing tomography.  In particular
we have quantified how the Figure of Merit for Dark Energy
parameters is affected by the choice of filters and observing
conditions.  We have generated catalogues from a range of proposed
surveys.  For fixed mock simulations and a fixed photo-z method
(ANNz) we explored the photo-z accuracy and systematics by varying
the set of filters and the magnitude limit of the surveys. The aim
of this was to compare the role of bands and to reduce biases given
that different mocks and photo-z methods may give different answers.

From the photometry, we find that there is an interplay between the
choice of filters (in particular the J, H and the u), the depth
(i.e. magnitude limit), and the removal of outliers based on the
photo-z errors. We find that if we are to get IR data from space the
improvements are greatly dependent on the quality of the ground data
available. For surveys that go down to the same magnitude
limit as the lensing survey in the RIZ band
the improvements that the J and H bands
bring are great. However, if we have ground based photometry,
particularly in the z and y bands, the improvements due to IR data
becomes smaller. There is a trade-off between the u and the IR,
meaning that the u band, provided it is deep enough, can play a
similar role as the IR data.

In summary, our main conclusions are:

\begin{itemize}

\item {The addition of J+H to griz+RIZ dramatically reduces the scatter in
individual photo-z, in particular for the shallow griz surveys.}

\item {The u band filter is effective in removing outliers and can play
the same role as the IR filters but only if the RIZ depth is
significantly larger than the depth of the lensing survey
chosen.}

\item{The results presented here depend on galaxy formation scenarios
which are encoded into the mock catalogues.
The main source for catastrophic photometric redshifts for an
optical survey arises from heavily reddened galaxies as
well as very blue starburst galaxies. It is hard to distinguish between a
higher redshift galaxy and a lower redshift reddened galaxy with
optical colours only below $z\sim1.2$. The opposite is true above $z\sim1.7$.
The inclusion of u band data or IR data breaks this degeneracy.}

\item{Our derived Figure of Merit $FOM = [\delta w_p \delta
w_a]^{-1} \sim 160$  can be obtained with a realistic mock
catalogue solely with weak lensing data. This would be a
significant increase over current estimates and also the next
generation of surveys. For comparison a current estimate of the
error on w is $[\delta w_p]^{-1} \sim 10$.}

\item{Given an ambitious ground based survey such as the LSST + IR bands
from space, only a marginal improvement in FOM can be achieved
by increasing the photometric accuracy of the visible bands
(i.e. the inclusion of deep u band photometry and deeper z and y
bands only increases the FOM fractionally from 156 to 168 (see
Table \ref{tbl:foms})). Since increasing the accuracy of the
visible bands will likely require deep space lensed data, the
extra cost of this is difficult to justify. This means that LSST
is the ideal counterpart to a DUNE like survey. However deeper
spaced based missions might have other ideal ground based
photometric matches.}

\item {The FOM improvement obtained from the addition of IR data depends on the quality of the
ground based optical data. The FOM is increased by a factor ranging from 1.7
down to 1.3 for realistic mock surveys.}

\item {The required number of spectroscopic redshifts needed depends on the
number of galaxies to train a neural network and also the
quantity of galaxies needed to calibrate the photometric redshifts.
We argue that a value
around $10^5$ in each redshift bin will be necessary for
weak lensing studies from
future space based missions.}

\item {The `cleaning' of outliers is effective. There is a trade-off
between reducing the photo-z error by removing galaxies,
but increasing the shot noise. It is possible to clean a photo-z catalogue
without decreasing the FOM significantly.
We conclude that
this is an effective way to decrease systematic effects
from a weak lensing survey and is an alternative to colour tomography.}

\end{itemize}

The general conclusion from our study is that combining weak lensing
measurements from space and photometric redshifts from
optical ground-based data is indeed a very attractive way to
constrain Dark Energy properties.
Furthermore, the use of IR from space can significantly improve
the accuracy of Dark Energy measurements by 30-70 percent.

\section*{Acknowledgements.}

We thank Manda Banerji, Sarah Bridle, Samuel Farrens, Huan Lin, Alex
Refregier, the DES and DUNE weak-lensing and photometric redshift
working groups and the COSMOS team for useful discussions. OL
acknowledges a PPARC Senior Research Fellowship. This research was
carried out in part at the Jet Propulsion Laboratory, California
Institute of Technology, under a contract with the National
Aeronautics and Space Administration and funded through the internal
Research and Technology Development program.

\bibliography{./reference/aamnem99,./reference/ref_data_base}

\end{document}